\documentclass[aps,prstab,superscriptaddress,twocolumn]{revtex4}
\usepackage[english]{babel}
\usepackage{graphicx}
\usepackage{booktabs}
\usepackage{amssymb}
\usepackage{graphics}
\usepackage{amsmath}
\usepackage{placeins}
\usepackage{hyperref}
\usepackage{epsfig}
\usepackage{verbatim}
\usepackage{threeparttable}

\begin{document}

\title{ Single shot cathode transverse momentum imaging in high brightness photoinjectors }

\author{Peng-Wei Huang}
\altaffiliation{ hpw17@mails.tsinghua.edu.cn}
\affiliation{ Deutsches Elektronen-Synchrotron DESY, Platanenallee 6, 15738 Zeuthen, Germany}
\affiliation{ Department of Engineering Physics, Tsinghua University, Beijing 100084,  China}

\author{Houjun Qian}
\altaffiliation{ houjun.qian@desy.de}
\affiliation{ Deutsches Elektronen-Synchrotron DESY, Platanenallee 6, 15738 Zeuthen, Germany}

\author{Ye Chen}
\affiliation{ Deutsches Elektronen-Synchrotron DESY, Platanenallee 6, 15738 Zeuthen, Germany}

\author{Daniele Filippetto}
\affiliation{ Lawrence Berkeley National Lab, Berkeley, CA 94704, USA}

\author{Matthias~Gross}
\affiliation{ Deutsches Elektronen-Synchrotron DESY, Platanenallee 6, 15738 Zeuthen, Germany}

\author{Igor~Isaev}
\affiliation{ Deutsches Elektronen-Synchrotron DESY, Platanenallee 6, 15738 Zeuthen, Germany}

\author{Christian Koschitzki}
\affiliation{ Deutsches Elektronen-Synchrotron DESY, Platanenallee 6, 15738 Zeuthen, Germany}

\author{Mikhail~Krasilnikov}
\affiliation{ Deutsches Elektronen-Synchrotron DESY, Platanenallee 6, 15738 Zeuthen, Germany}

\author{Shankar Lal}
\affiliation{ Deutsches Elektronen-Synchrotron DESY, Platanenallee 6, 15738 Zeuthen, Germany}

\author{Xiangkun Li}
\affiliation{ Deutsches Elektronen-Synchrotron DESY, Platanenallee 6, 15738 Zeuthen, Germany}

\author{Osip~Lishilin}
\affiliation{ Deutsches Elektronen-Synchrotron DESY, Platanenallee 6, 15738 Zeuthen, Germany}

\author{David Melkumyan}
\affiliation{ Deutsches Elektronen-Synchrotron DESY, Platanenallee 6, 15738 Zeuthen, Germany}

\author{Raffael Niemczyk}
\affiliation{ Deutsches Elektronen-Synchrotron DESY, Platanenallee 6, 15738 Zeuthen, Germany}

\author{Anne Oppelt}
\affiliation{ Deutsches Elektronen-Synchrotron DESY, Platanenallee 6, 15738 Zeuthen, Germany}

\author{Fernando Sannibale}
\affiliation{ Lawrence Berkeley National Lab, Berkeley, CA 94704, USA}

\author{Hamed Shaker}
\affiliation{ Deutsches Elektronen-Synchrotron DESY, Platanenallee 6, 15738 Zeuthen, Germany}

\author{Guan Shu}
\affiliation{ Deutsches Elektronen-Synchrotron DESY, Platanenallee 6, 15738 Zeuthen, Germany}

\author{Frank Stephan}
\affiliation{ Deutsches Elektronen-Synchrotron DESY, Platanenallee 6, 15738 Zeuthen, Germany}

\author{Chuanxiang Tang}
\affiliation{ Department of Engineering Physics, Tsinghua University, Beijing 100084,  China}

\author{Grygorii Vashchenko}
\affiliation{ Deutsches Elektronen-Synchrotron DESY, Platanenallee 6, 15738 Zeuthen, Germany}

\author{Weishi Wan}
\affiliation{ School of Physical Science and Technology, Shanghai Tech University, Shanghai 201210, China}

\date{\today}
\begin{abstract}
  In state of the art photoinjector electron sources, thermal emittance from photoemission dominates the final injector emittance. Therefore, low thermal emittance cathode developments and diagnostics are very important. Conventional thermal emittance measurements for the high gradient gun are time-consuming and thus thermal emittance is not measured as frequently as quantum efficiency during the lifetime of photocathodes, although both are important properties for the photoinjector optimizations. In this paper, a single shot measurement of photoemission transverse momentum, i.e., thermal emittance per rms laser spot size, is proposed for photocathode RF guns. By tuning the gun solenoid focusing, the electrons' transverse momenta at the cathode are imaged to a downstream screen, which enables a single shot measurement of both the rms value and the detailed spectra of the photoelectrons' transverse momenta. Both simulations and proof of principle experiments are reported.
\end{abstract}

\maketitle

\section{Introduction}

Modern linear accelerator based facilities, such as x-ray free electron lasers (XFEL), ultrafast electron diffration/microscopy (UED/UEM) enable the exploration of ultrafast dynamics at the atomic level \cite{lessner2016report}. Both types of facilities need high brightness electron sources, which define the lower limit of the normalized emittance at the linac exit. For a given linac energy, a smaller normalized emittance improves the beam brightness and hence the XFEL gain and the pulse energy. Normalized emittance as low as 0.1~mm~mrad for a 100~pC beam is wished for the next generation XFELs \cite{lessner2016report,musumeci2018advances}. In state of the art photoinjectors, thermal emittance dominates the final injector emittance, so low thermal emittance cathode developments and characterizations are extremely important for further emittance reduction \cite{dowell2010cathode,pitzpaper,cedominate}. In the real gun environment, cathode properties change over time due to residual gases in the gun vacuum, particle back bombardments, RF heating and radiations \cite{Charge2013,le2013quantum,Filippetto2015Cesium,Jones2017Evolution,Huang2019Photoemission}. Both quantum efficiency (QE) and thermal emittance should be routinely measured to help injector tuning and optimizations. In practice, the conventional thermal emittance measurement is time consuming, thus it is not done as frequently as QE measurement of the cathode in the gun.

Generally, the transverse position and transverse momentum of the photoemission electrons are not correlated and the thermal emittance $\varepsilon_{th}$ at the cathode can be calculated by
\begin{eqnarray}
  \varepsilon_{th} = \sigma_{x0} \dfrac{\sigma_{p_{x0}}}{m_0c}.
\end{eqnarray}
The transverse size $\sigma_{x0}$ is the same as the laser spot size, and the normalized rms transverse momentum, the ratio between the rms transverse momentum $\sigma_{p_{x0}}$ and the product of electron mass $m_0$ and the speed of light $c$, is a key property of the cathode. For most of the working conditions, the rms transverse momentum can be fitted through thermal emittance measurements with different laser spot sizes.

Conventional methods to measure thermal emittance include gun solenoid scan, quadrupole scan and pepper-pot mask \cite{lee2015review}. Cathode laser spot sizes can be measured on the virtual cathode plane with a camera. However, a complete measurement to fit the rms cathode transverse momentum takes many data points, therefore is time consuming and vulnerable to machine stability and measurement errors. Besides, the long measurement time makes such techniques not suitable for routine measurements of cathode transverse momentum.

A single shot method for measuring the transverse electron momentum distribution from a photocathode was first developed using a low energy cathode diagnostic chamber at Lawrence Berkeley National Laboratory (LBNL) \cite{feng2015novel}. The electrons get accelerated by a static electric field and drift freely before hitting a screen. The cathode laser spot size is reduced so that the beam size on the screen is dominated by the transverse momentum spread from photoemission. For keV level DC acceleration, this corresponds to a sub mm laser spot size. For MeV level RF guns, the cathode laser spot size requirement is expected to reduce dramatically due to higher longitudinal beam momentum, so such a free expansion technique is more suitable for low energy DC guns, in which the cathode field is usually below 10 MV/m.

Another single shot method for measuring the photoemission transverse momentum distribution was developed at University of Illinois \cite{Berger2012excited}. A pair of solenoids after a 20 kV DC gun form a focal plane on a downstream screen, so the beam size on the focal plane is linearly proportional to the cathode transverse momentum. Such a method does not require a small laser spot size at the cathode, but it needs a solenoid after the gun and a proper focusing tuning. Such a technique can be easily adapted to high gradient RF photoinjectors, where a gun solenoid is naturally located at the gun exit.



In this paper, the first single shot measurement of cathode transverse momentum spread in a high gradient RF gun is presented. The basic idea of the proposed method and some analytical derivations are introduced first. Then a confirmation of this concept by simulation is provided. The systematic errors of the method are discussed. Finally, proof of principle experiments carried out in a real accelerator environment, the Photo Injector Test Facility at DESY, Zeuthen site (PITZ), are presented.

\section{Basic idea}

\begin{figure}[!tbp]
   \includegraphics[width=8cm, height=5cm]{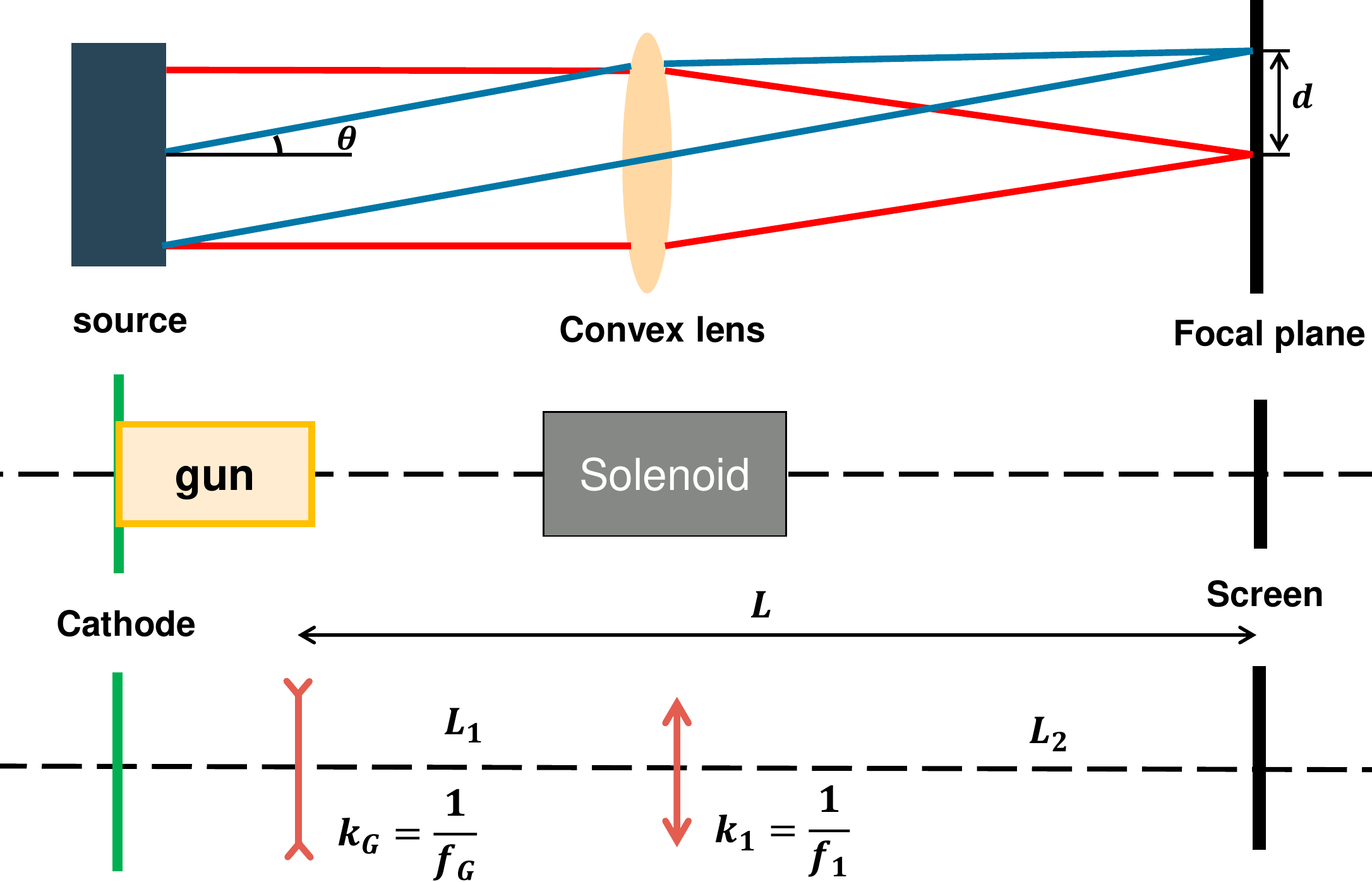}
   \caption{ The basic concept of the cathode transverse momentum imaging. $\theta$ is the angle between the light trajectory and horizontal line. $d$ is the distance between the hitting point and image center. A simplified electron beam line is set up. The first element is the RF gun with a defocusing lens of strength $k_G$. After a drift of distance $L_1$, the second element is a focusing lens with $k_1$. Then comes another drift $L_2$ and the final element is the screen. $L$ is the sum of $L_1$ and $L_2$.}\label{theory}
\end{figure}

In optics, it is well known that parallel light rays can be focused by a convex lens into one point at the focal plane and the distance to the central axis, $d$, is linearly proportional to the initial direction $\theta$, shown in Fig.\;\ref{theory}. Therefore, the transverse angle distribution of the source is imaged onto the focal plane. Electron optics with negligible space charge effects are very similar to light optics, so the idea can be adapted to measure the transverse momentum from the electron source. A solenoid magnet focuses the electron beam in a similar way as a convex lens does with light rays, except for the Larmor frame rotation. Consequently, a proper tuning of the solenoid is able to image the electron's initial transverse momentum onto the downstream observation screen. In accelerator language, the imaging condition is met when the first element of the transport matrix from cathode to the observation screen is zero:

\begin{eqnarray}\label{image_con}       
\left(                 
   \begin{array}{c}
    x \\
    \frac{p_x}{m_0c} \\  
  \end{array}
\right)  \ = \
\left(                 
   \begin{array}{cc}
    0 & M_{12} \\
    M_{21} & M_{22} \\  
  \end{array}
\right)
\left(                 
   \begin{array}{c}
    x_0 \\
    \frac{p_{x0}}{m_0c} \\  
  \end{array}
\right).
\end{eqnarray}
Here the matrix works in the Larmor rotation coordinate, decoupling the $x$ and $y$ planes. At the imaging condition, a linear relationship is established between the electron's transverse momentum at the cathode and the corresponding transverse position on the observation screen. A mathematical expression for calculating the normalized cathode transverse momentum is derived as
\begin{eqnarray}\label{cal_emit}
  \dfrac{\varepsilon_{th}}{\sigma_{x0}} = \dfrac{\sigma_{p_{x0}}}{m_0c} = \dfrac{\sigma_x(L)}{M_{12}},
\end{eqnarray}
where $\sigma_x(L)$ is the rms spot sizes at the screen. Once the imaging condition is found, $M_{12}$ belongs to the transport matrix between the cathode and the screen. Only a single shot is needed to obtain $\sigma_x(L)$, the spot size at the screen, and the normalized cathode transverse momentum can be calculated with $M_{12}$.

For the proposed beam line in Fig.\;\ref{theory}, an analytical transport matrix can be derived with some simplifications. The transverse RF focusing from cathode to gun exit is simplified to be a defocusing lens following reference \cite{kim1989rf}, and the defocusing strength $k_G$ is related to the beam momentum and gun field profile. The beam size change in the gun is neglected in this simplified model under the zero space charge limit. The solenoid is simplified to be a thin lens with strength $k_1$, and its Larmor rotation is neglected in Eq.\;(\ref{matrix}) thanks to the rotational symmetry of the beamline. $L_1$ is the distance between the two lenses, i.e. the distance from the gun exit iris to gun solenoid center, and $L_2$ is the distance from the solenoid center to the screen. With these simplifications, the transport matrix can be derived as

\begin{eqnarray}\label{matrix}       
\left(                 
   \begin{array}{c}
    x \\
    x' \\  
  \end{array}
\right)  &=&
\left(                 
   \begin{array}{cc}
    1 & L_2 \\
    0 & 1 \\  
  \end{array}
\right)
\left(                 
   \begin{array}{cc}
    1 & 0 \\
    -k_1 & 1 \\  
  \end{array}
\right)
\left(                 
   \begin{array}{cc}
    1 & L_1 \\
    0 & 1 \\  
  \end{array}
\right)
\left(                 
   \begin{array}{cc}
    1 & 0 \\
    k_G & 1 \\  
  \end{array}
\right)
\left(                 
   \begin{array}{c}
    x_0 \\
    x'_0 \\  
  \end{array}
\right) \nonumber  \\
&=&
\left(                 
   \begin{array}{cc}
    1+k_GL-(1+k_GL_1)L_2k_1 & L-L_1L_2k_1 \\
    k_G-(1+k_GL_1)k_1 & 1-L_1k_1 \\  
  \end{array}
\right)
\left(                 
   \begin{array}{c}
    x_0 \\
    x'_0 \\  
  \end{array}
\right), \nonumber \\
\end{eqnarray}
where $L$ is the sum of $L_1$ and $L_2$. The expression for $x'_0$ is
\begin{eqnarray}
  x'_0 &=& \frac{p_{x0}}{p_z},
\end{eqnarray}
where $p_z$ is the longitudinal beam momentum at gun exit. The condition in Eq.\;(\ref{image_con}) is met when the solenoid strength $k_1$ is chosen as
\begin{eqnarray}
  k_1 &=& \dfrac{1+k_GL}{(1+k_GL_1)L_2}.
\end{eqnarray}
The gun defocusing strength $k_G$ can be estimated through \cite{kim1989rf}
\begin{eqnarray}
  k_G = \dfrac{\text{d} x'}{\text{d} x} = \dfrac{1}{p_z}\dfrac{\text{d} p_x}{\text{d} x} = \dfrac{eE_0}{2cp_z}\sin{\phi},
\end{eqnarray}
where $E_0$ is the gun's peak electric field and $\phi$ is the synchronous phase of electron beam at gun exit \cite{kim1989rf}. With the above derivation, $M_{12}$ in Eq. (\ref{image_con}) is
\begin{eqnarray}
  M_{12} &=& (L-L_1L_2k_1)\dfrac{m_0c}{p_z}=\dfrac{L_2}{1+k_GL_1}\dfrac{m_0c}{p_z}. \label{m12}
\end{eqnarray}

Two messages can be obtained from the derivation. First, a proper setting of the solenoid is analytically confirmed to satisfy the imaging condition. Second, $L_2$ should be large enough to achieve a reasonable beam size at the focal plane screen. The beam size cannot be too small due to the spatial resolution of the imaging system. In addition, since very low bunch charge is required to minimize the space charge in the measurement, a too large beam size is not preferred due to the signal to noise ratio (SNR) issues. The above analytical model is simplified, providing a rough and fast estimation for $M_{12}$. The exact value of $M_{12}$ will be obtained from ASTRA simulations \cite{astra} and cross-checked with conventional thermal emittance measurements, which will be discussed later.

\section{Simulation proof}

Simulations based on the PITZ beam line have been done to confirm the feasibility of the method. The assumed beam line includes the PITZ gun, the gun solenoid and a screen as shown in Fig.\,\ref{theory}. In simulation, the initial electron beam is assumed to be transversely uniform (rms size 0.25 mm) and longitudinally Gaussian (7 ps FWHM), following the laser distribution at PITZ. The peak electric field is selected to be $E_z =$ 53 MV/m, which is lower than the nominal value due to the concern of dark current for low charge beam measurement, and the corresponding final electron momentum is 5.8 MeV/c at the maximum mean momentum gain (MMMG) phase of the gun. The beam was tracked to the position z = 5.28 m from the cathode, where a high efficiency LYSO screen \cite{niemczyk2019comparison} is used for low charge beam size measurements.

\begin{figure}[!tbp]
   \includegraphics*[width=.95\columnwidth]{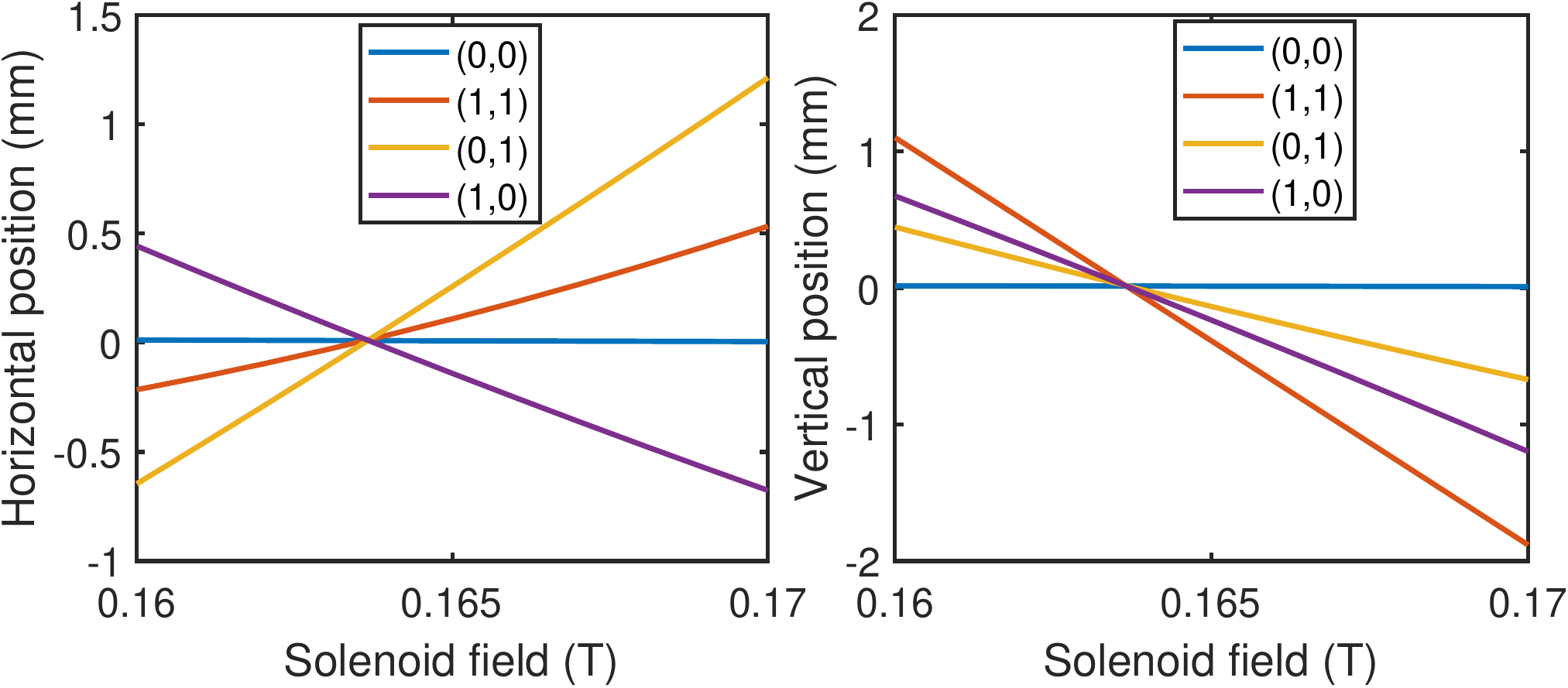}
   \caption{ The variation of beam position on the observation screen versus solenoid field is simulated by the ASTRA code. The legend contains the particles' initial transverse position on the cathode with a unit of mm. }\label{beamoffset}
\end{figure}

The simulation verification of the imaging condition by tuning the solenoid strength is done with electron beams starting from several different positions on the cathode. When $M_{11}$ is zero, the electron beams should end at the same position on the observation screen. The simulation results, presented in Fig.\;\ref{beamoffset}, show that beams from different positions on the cathode overlap with each other at the same solenoid setting, indicating that the imaging condition $M_{11}=0$ has been met with solenoid tuning.

\begin{figure}[!tbp]
   \includegraphics*[width=.8\columnwidth]{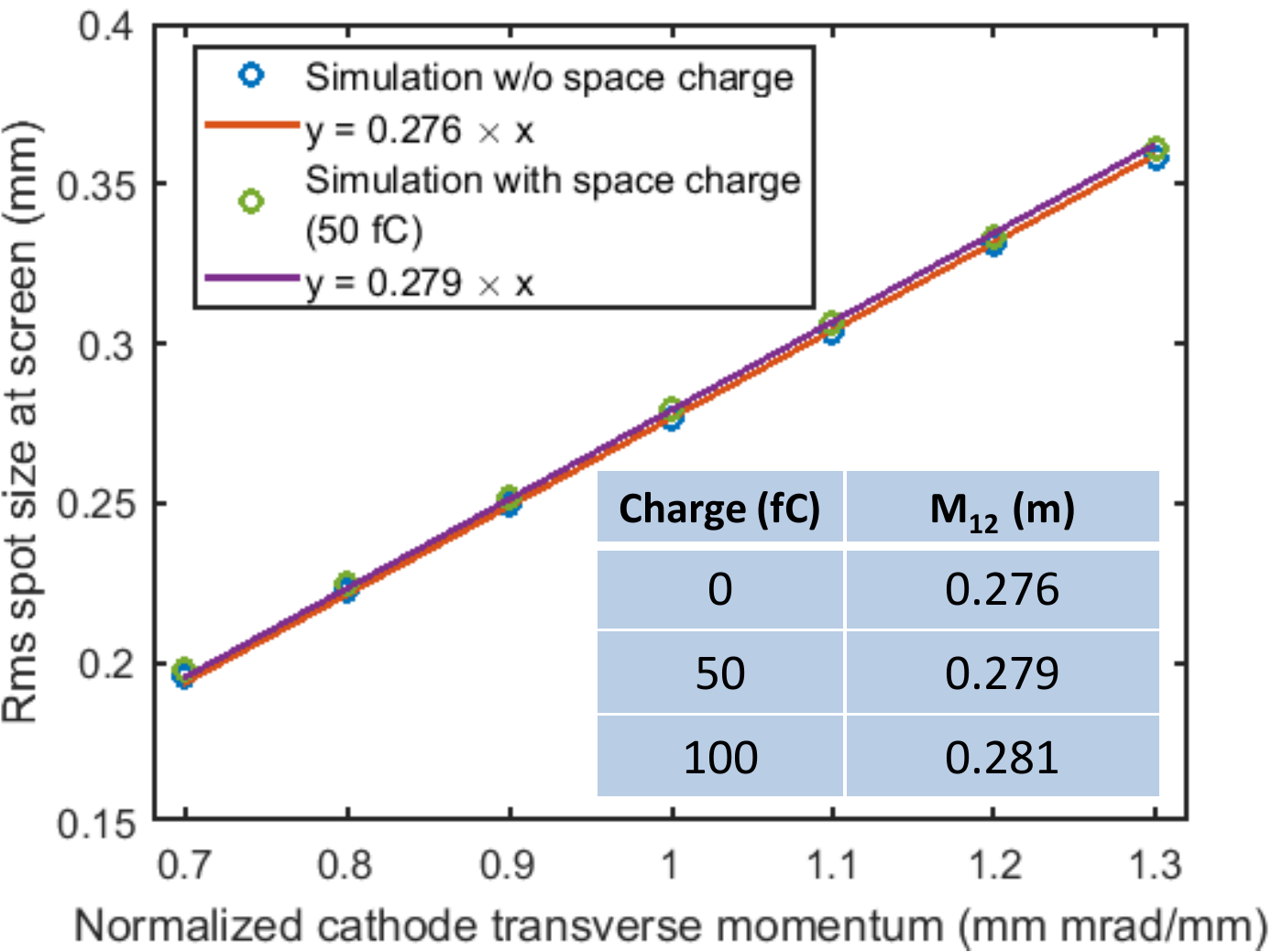}
   \caption{ The beam size at the observation screen is linearly proportional to the cathode transverse momentum in simulation with and without space charge. The fitting formula is presented in the legend. Here the solenoid has been set to meet the imaging condition.}\label{size_emit}
\end{figure}

 At $M_{11}=0$, the linear relation between the rms cathode transverse momentum and the corresponding beam size at the screen is also examined. The simulation results are presented in Fig.\;\ref{size_emit}. Good agreement is achieved with a linear fit and it is a support to Eq.\;(\ref{cal_emit}). The slope of the fitting line is the value of $M_{12}$. When the charge is below 100 fC, the space charge effect on $M_{12}$ is negligible, according to the simulations. The analytical expression of Eq.\;(\ref{m12}) gives the value of $M_{12}$ as 0.3 m, which is reasonably close to the simulation result.

\begin{figure}[!tbp]
   \includegraphics*[width=.95\columnwidth]{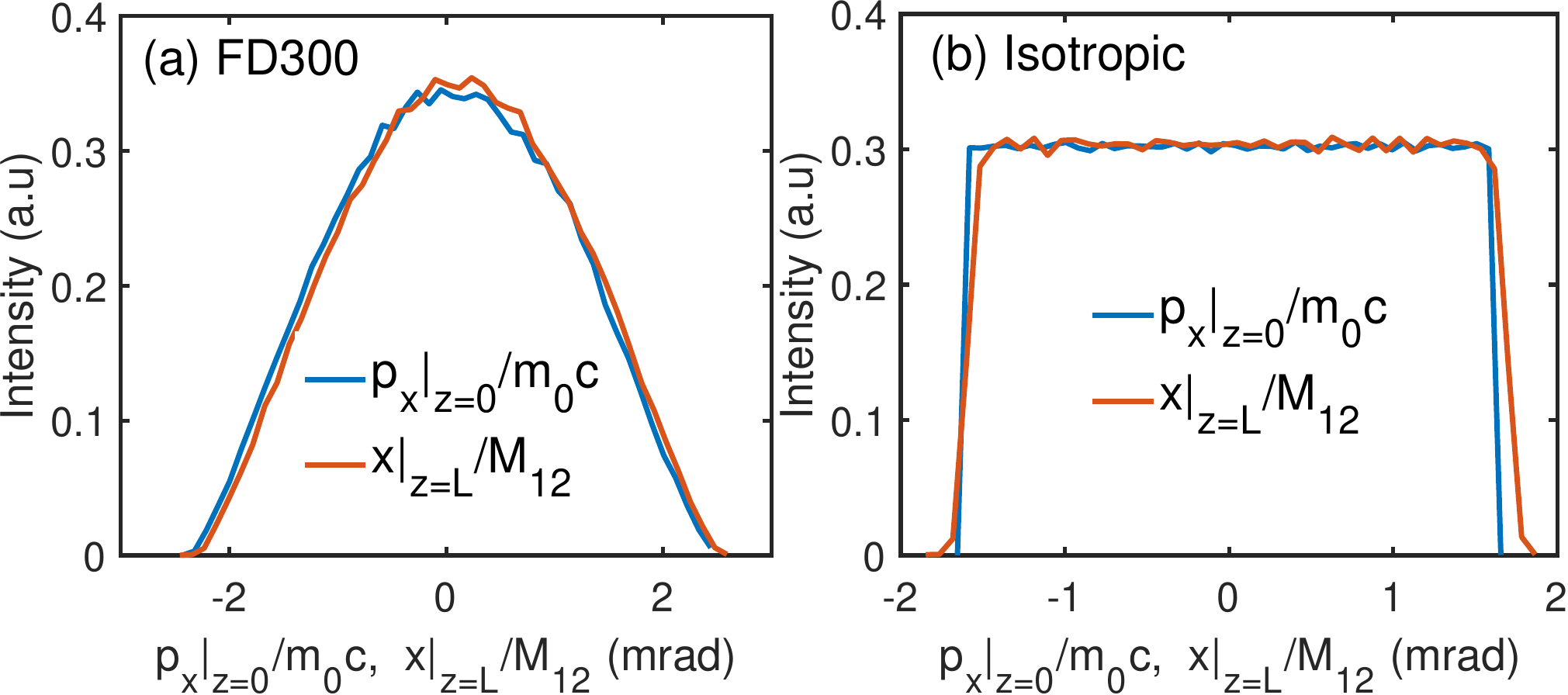}
   \caption{ A comparison between the initial transverse momentum distribution and the transverse projection of the image at the screen. The FD300 model and isotropic model in ASTRA simulation are used as examples to generate electrons with different momentum distribution, shown in (a) and (b) respectively. }\label{mom_ima}
\end{figure}

 The momentum imaging is also verified with different initial electrons' momentum distributions. The initial transverse momentum $p_x|_{z=0}$ is linearly proportional to the final transverse position $x|_{z=L}$, so
\begin{eqnarray}
  F(\dfrac{p_x|_{z=0}}{m_0c}) &=& F(\dfrac{x|_{z=L}}{M_{12}}),
\end{eqnarray}
where $F$ is a probability density function. Simulations with initial electron distributions generated from two different emission models in ASTRA, FD300 and isotropic, are presented in Fig.\;\ref{mom_ima}. The consistency between the cathode transverse momentum distribution and the beam projection at the screen is observed for both emission models. Measurement of the momentum distribution in addition to the rms value would be helpful to establish a reliable photoemission model.

\section{systematic errors}






\begin{figure}[!tbp]
   \includegraphics*[width=.8\columnwidth]{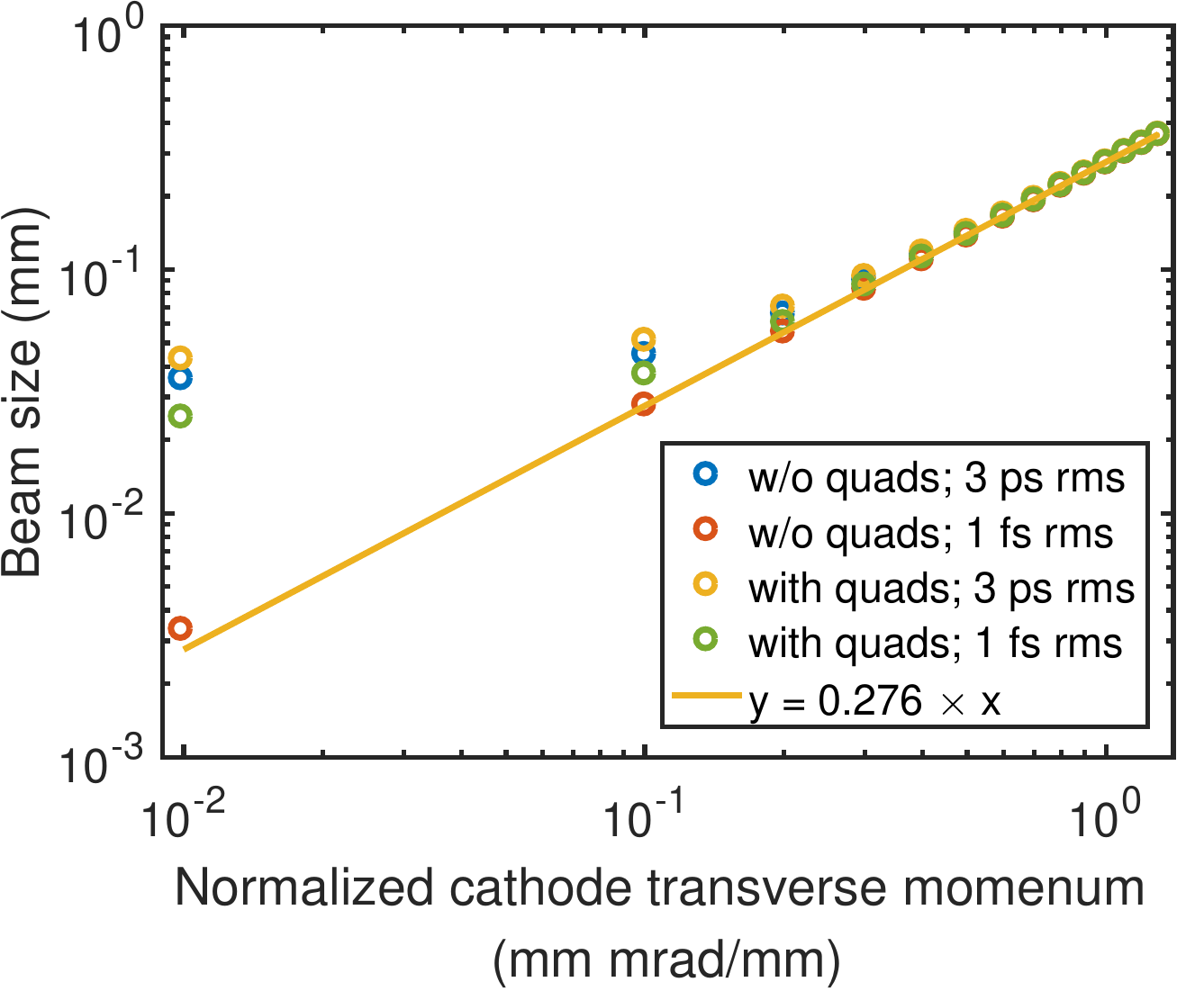}
   \caption{ The beam size at the observation screen versus the normalized cathode transverse momentum at MMMG phase. The circles are the simulation results. The blue and red circles assume an ideal solenoid without quadrupole field errors and the bunch lengths are 3 ps rms and 1 fs rms, respectively. The yellow and green circles include the quadrupole field errors of the solenoid. The line is the linear fitting. }\label{size_emit_0}
\end{figure}

\begin{figure}[!tbp]
   \includegraphics*[width=.8\columnwidth]{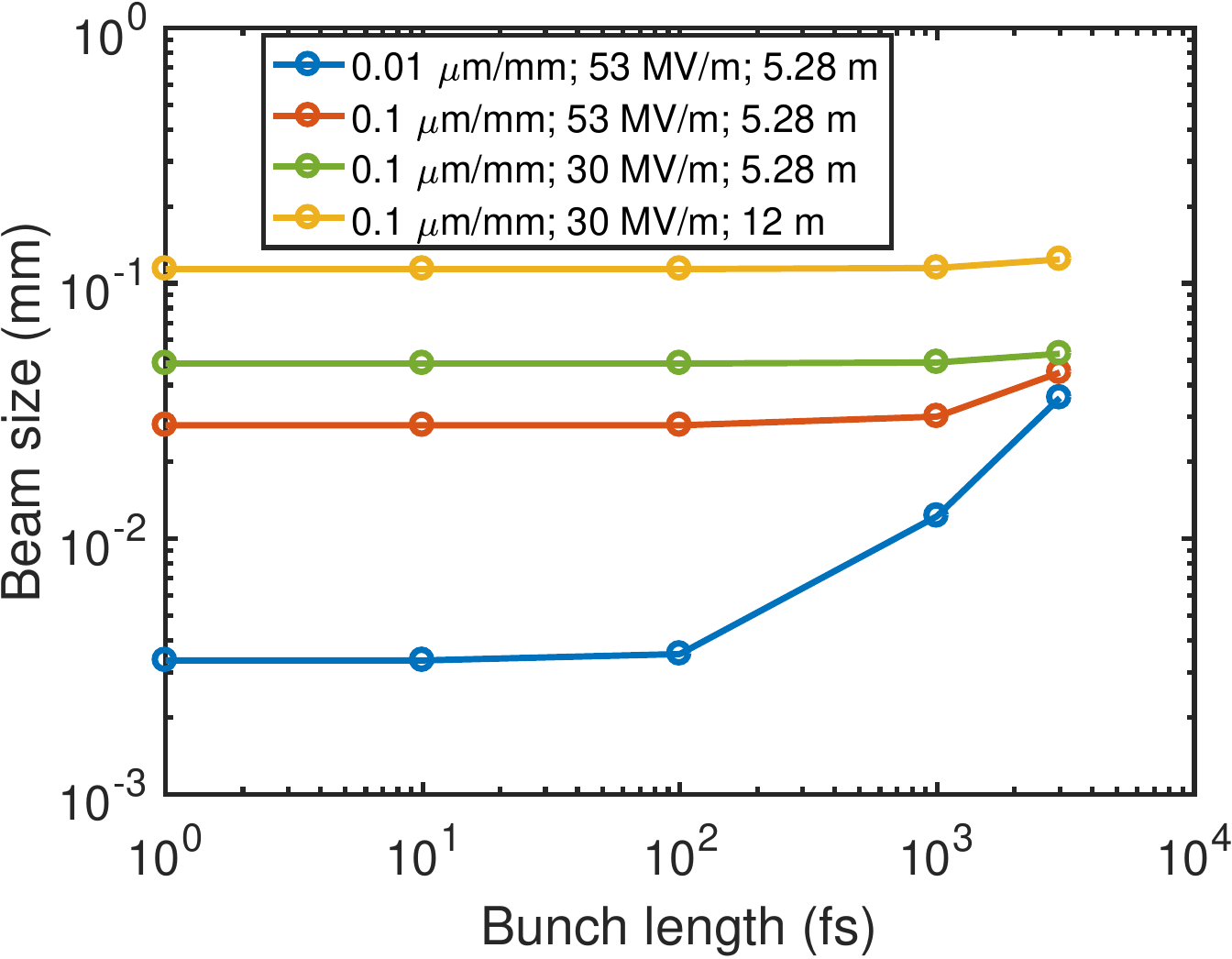}
   \caption{ The beam size evolution with the bunch length. The items in the legend are the normalized cathode transverse momentum, the peak cathode field and the distance from the cathode to the screen, respectively. }\label{size_vs_length_phase0}
\end{figure}

The proposed method is robust for the measurements of high cathode transverse momentum, like 1 mm mrad/mm. However, for cathodes with extremely low cathode transverse momentum, the measurement is vulnerable to many factors, like RF induced emittance growth and quadrupole field error in the solenoid. The following discussions are based on simulations without space charge.

The RF induced emittance growth is due to the time dependent RF focusing in the gun, and its expression is \cite{kim1989rf}
\begin{eqnarray}\label{RF_emit}
  \varepsilon_{RF} &=& \dfrac{eE_z}{2m_0c^2}\sigma_x^2\sigma_{\phi}\cos{\phi_f}
\end{eqnarray}
where $\sigma_x$ is the rms beam size and $\sigma_{\phi}$ is the rms bunch length in terms of RF phase. $e$ is the elementary charge and $\phi_f$ is the exit phase. Simulations with the same setup as for Fig.\;\ref{size_emit} have been done for electron beams with different normalized cathode transverse momentum from 0.01 to 1.3 mm mrad/mm, covering most of the interesting photocathodes. As shown in Fig.\;\ref{size_emit_0}, the beam size at the screen starts to deviate from the linear expectation when the normalized cathode transverse momentum is below 0.2 mm mrad/mm with a bunch length of 3 ps rms. Another simulation with a much shorter beam ($\sim$ 1 fs) almost removes the deviations even at 0.01 mm mrad/mm, confirming the error comes from RF emittance. It is more meaningful to determine the bunch length limit within an acceptable measurement error. The evolution of the beam size with bunch length is presented in  Fig.\;\ref{size_vs_length_phase0}. For 0.01 mm mrad/mm, the beam size reaches a constant when the bunch length is below 100 fs, while the limit can be extended to 1 ps rms for 0.1 mm mrad/mm. Since the thermal limit is 0.22 mm mrad/mm at room temperature \cite{feng2015thermal}, a bunch length of 1 ps rms is short enough for most of the measurements. According to Eq. (\ref{RF_emit}), a lower gun gradient can also reduce the RF emittance. When the peak field is 30 MV/m, a bunch length of 3 ps rms is short enough enough for a reliable measurement of 0.1 mm mrad/mm. Besides, the beam size is larger. If the screen is moved further downstream the beamline, like z = 12 m, the beam size can be increased from $\sim$50 um to $\sim$100 um for better measurement accuracy at 0.1 mm mrad/mm. An optimization of both peak field and the screen position can result in a reasonable beam size for the target range of the normalized cathode transverse momentum.

\begin{figure}[!tbp]
   \includegraphics*[width=.8\columnwidth]{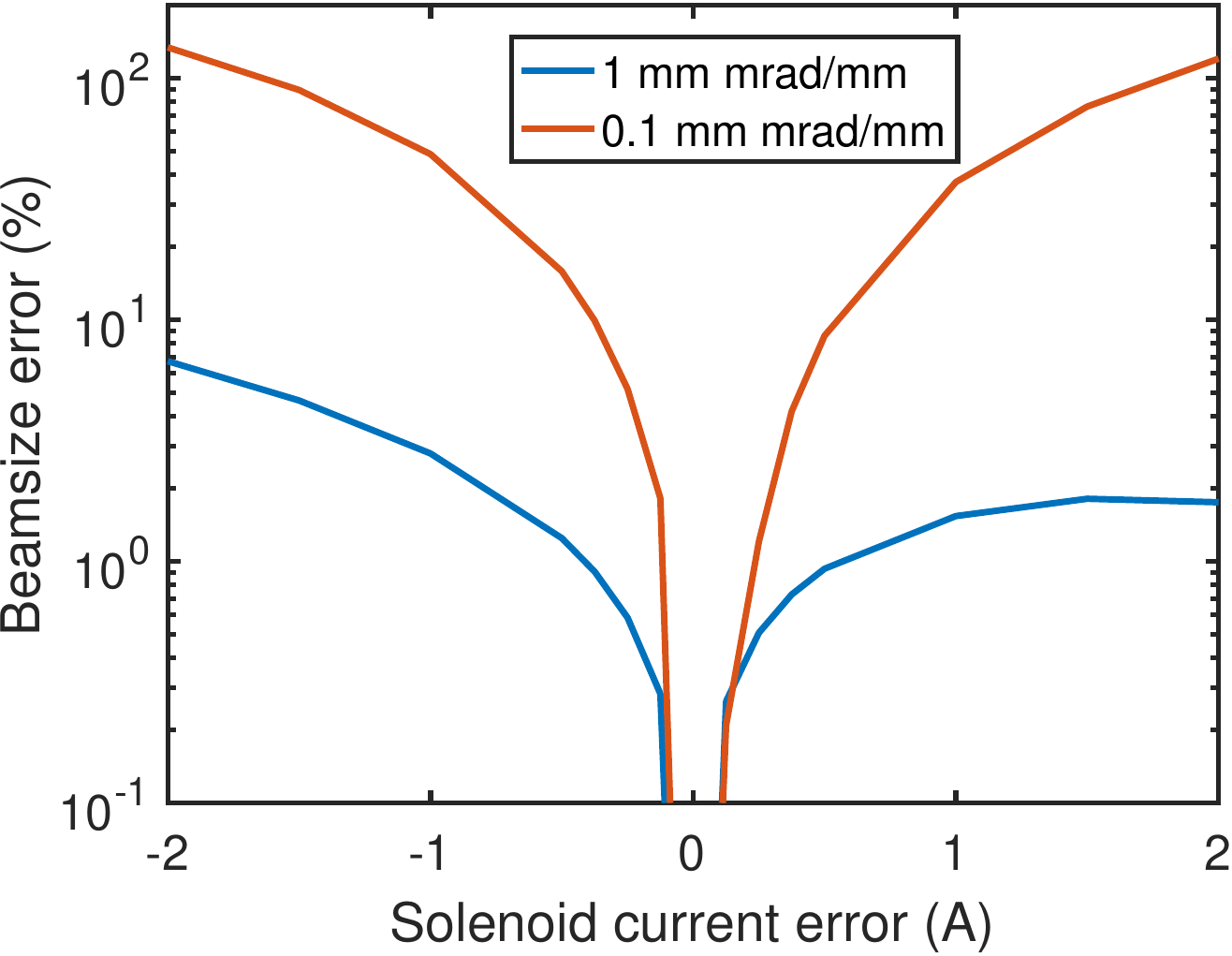}
   \caption{ The beam size error caused by the solenoid current error with respect to the momentum imaging condition. }\label{sol_cur_error}
\end{figure}

The solenoid is required in the proposed method to provide focusing strength. Any imperfections of the solenoid will exert influence on the measurements. The quadrupole field error in the solenoid is an important source of error and it will induce coupled transverse dynamics and result in emittance growth \cite{Krasilnikov,Dowell2018solenoid,Zheng2018overestimation,Zheng2019Experimental}. In the search of the solenoid current to meet the imaging condition, the existence of the quadrupole field will cause a deviation between the horizontal and vertical plane. At PITZ, the relative difference for solenoid current between X and Y plane is below 0.7 $\%$ and the influence on the measurement is presented in Fig.\;\ref{size_emit_0}. No significant effect is observed for the high normalized cathode transverse momentum measurement, while large errors can be found in the lower region. Fig.\;\ref{size_emit_0} also shows that the errors can not be eliminated by shortening the bunch length, indicating that the emittance growth brought by the quadrupole field error accounts for a dominant part. Quadrupole field correction is necessary for the measurements of the extremely low cathode transverse momentum.

In the experiment, the step size of the solenoid scan limits the accuracy of the target solenoid current within half of the step size. If the step size is 1 A, which is 0.3 $\%$ of the solenoid current for imaging in our example, the corresponding error is presented in Fig.\;\ref{sol_cur_error}. If the normalized cathode transverse momentum is 1 mm mrad/mm, the error is below 1.3 $\%$. However, when the normalized cathode transverse momentum is as low as 0.1 mm mrad/mm, a smaller step size is necessary, like 0.25 A, to obtain a small error of 1.8 $\%$.

\section{Proof of Principle Experiment}

\begin{figure}[!tbp]
   \includegraphics*[width=.8\columnwidth]{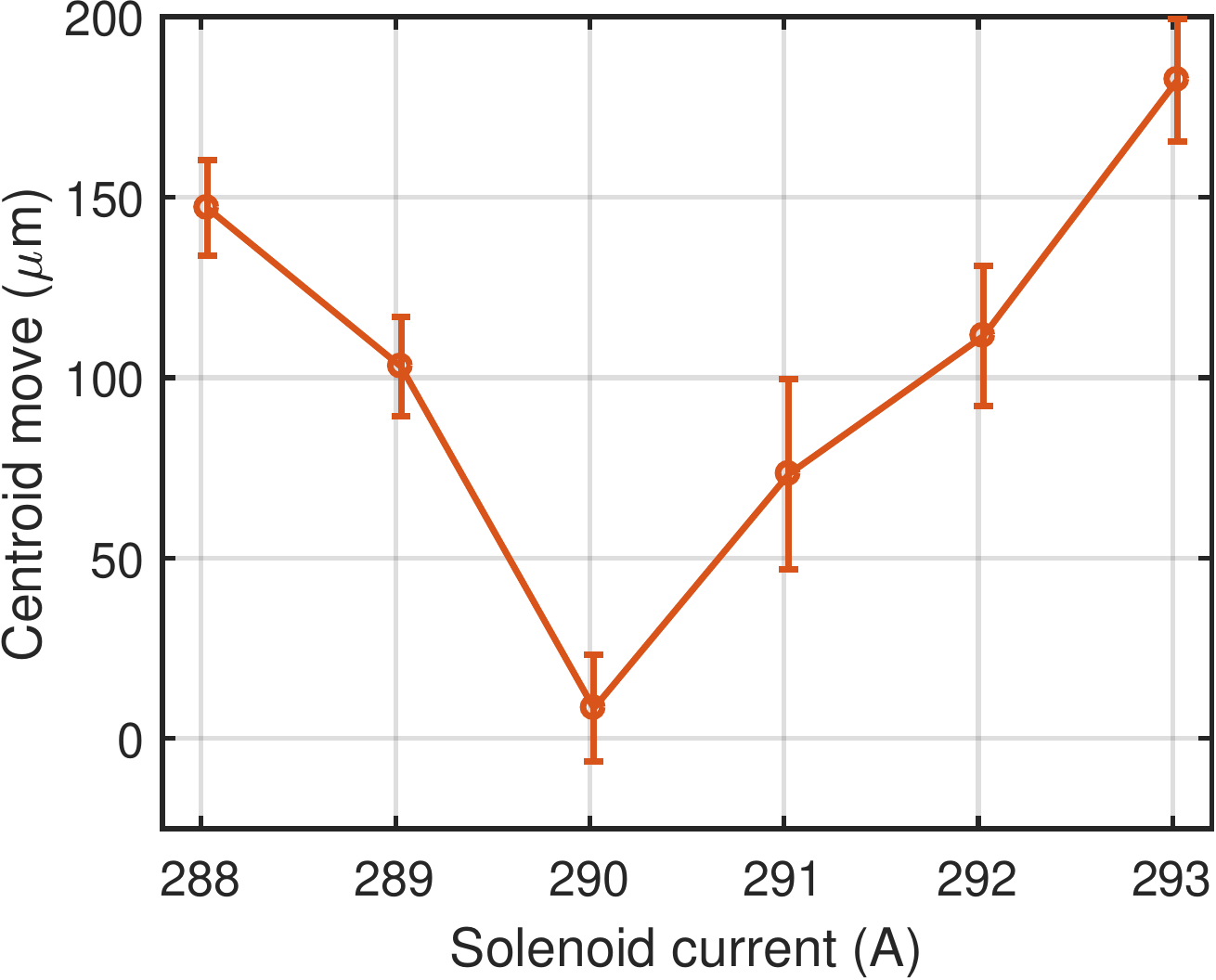}
   \caption{ One example of beam movement at the observation screen when the laser spot is moved by 0.5 mm vertically at the cathode in experiment.}\label{find_sol}
\end{figure}

Proof of principle experiments have been carried out at PITZ. The applied laser wavelength is 257 nm. The experimental condition is the same as mentioned in the simulation section. The first step of the measurements is looking for the solenoid current to meet the imaging condition where $M_{11}$ is zero. $M_{11}$ is measured by trajectory response test. The laser position is changed on the cathode and the corresponding beam movement is measured on the observation screen. The measurements repeat for a range of solenoid currents until the least centroid movement is found. An example is shown in Fig. \ref{find_sol}. The centroid movement can not reach exactly zero for several reasons. First, the solenoid scan step size is not infinitely small. Second, there are multipole field errors in the solenoid, which will make $M_{11}$ zero at different currents for $x$ and $y$ planes. Third, jitters exist in the machine. When the laser position is moved by 500 $\mu$m and the beam centroid movement at screen is below 50 $\mu$m, like the example in Fig.\;\ref{find_sol}, the term, $M_{11}$, is less than 0.1. At this case, the derivation of the mean square of beam size is expressed as
\begin{eqnarray}
  \langle x^2\rangle &=& M^2_{11}\langle x^2_0\rangle + M^2_{12}\langle (\frac{p_{x0}}{m_0c})^2\rangle.
\end{eqnarray}
With an initial rms beam size of 0.25 mm and $M_{12}$ of 0.276 m, if the target normalized cathode transverse momentum is 1~mm mrad/mm, the residual $M_{11}$ would contribute to a negligible ($<$\;0.5\;$\%$) overestimation of the results in our measurement. It should be noted that a large $M_{12}$ is good for the alleviation of the influence from the residual $M_{11}$. In the experiments, we found that it is not difficult to keep the least centroid movement below 50 $\mu$m with a step size 1 A for the solenoid current.

\begin{figure}[!tbp]
   \includegraphics*[width=.8\columnwidth]{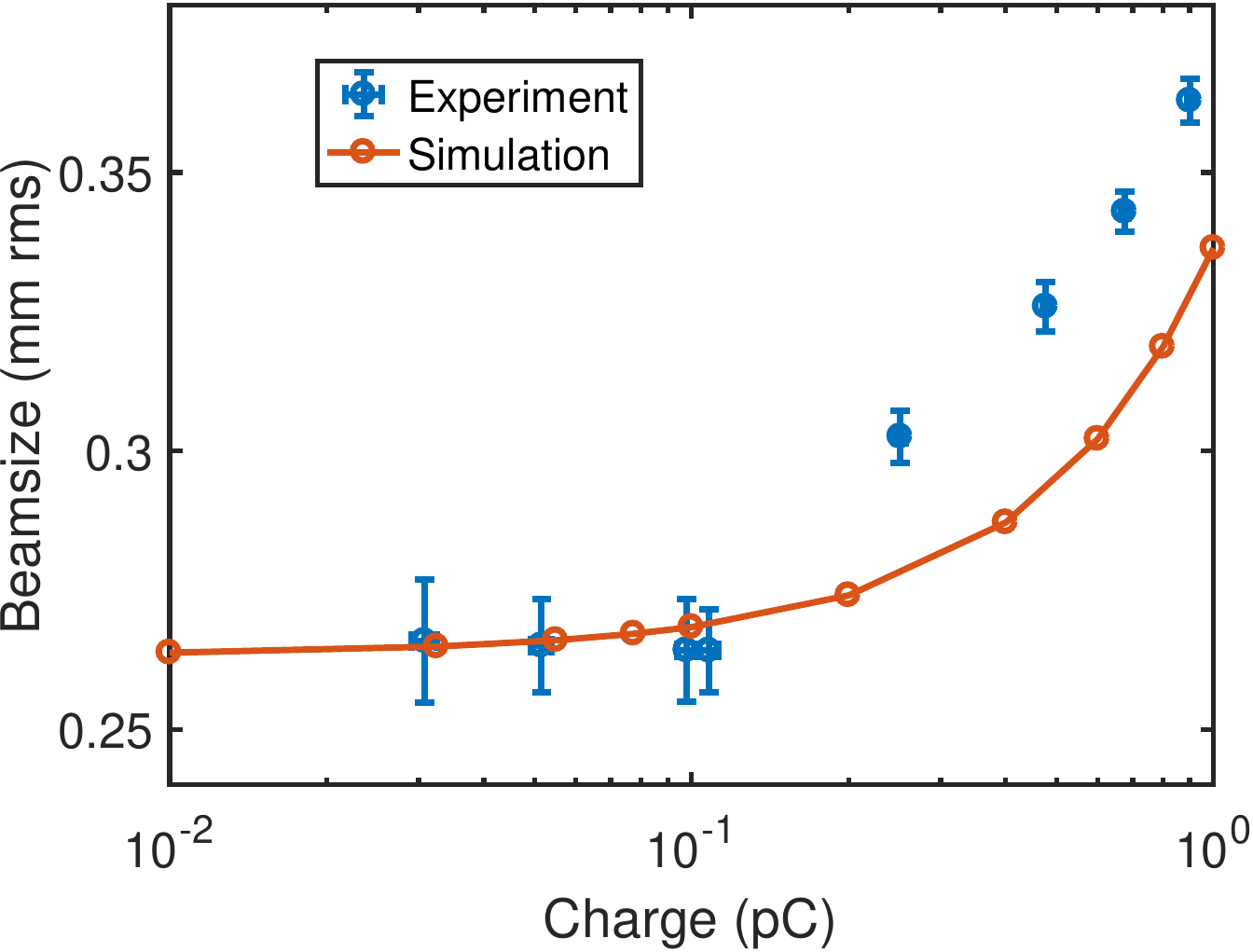}
   \caption{ The beam size variation with charge at the observation screen both in experiment and in simulation are presented. The laser diameter is 1 mm and the length is around 3 ps rms. The solenoid current is set to the imaging condition.}\label{space_charge}
\end{figure}

Space charge is one of the important reasons to cause the overestimation of the thermal emittance. In principle, the effect can be alleviated by low charge operation. However, extremely low charge beams are not favorable for the concern of the SNR of the image on the screen, increasing the errors in the image processing. Therefore, a compromise should be made between the space charge effect and SNR. The upper limit for negligible space charge effect should be determined experimentally. It is found that the beam size at the momentum imaging condition remains constant when the charge is below 100 fC, which is consistent with the simulation results, shown in Fig.\;\ref{space_charge}. Therefore 100 fC is the upper charge limit where space charge effect is negligible. The bunch charge was kept roughly at 50 fC in the following measurements.


\begin{figure}[!tbp]
   \includegraphics*[width=.8\columnwidth]{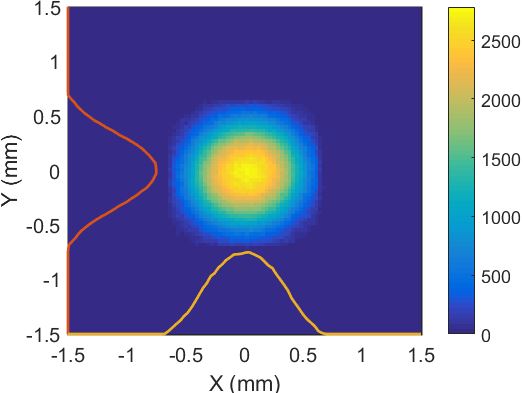}
   \caption{ A typical image in the momentum imaging measurements. }\label{typical_pic}
\end{figure}

\begin{figure}[!tbp]
   \includegraphics*[width=.8\columnwidth]{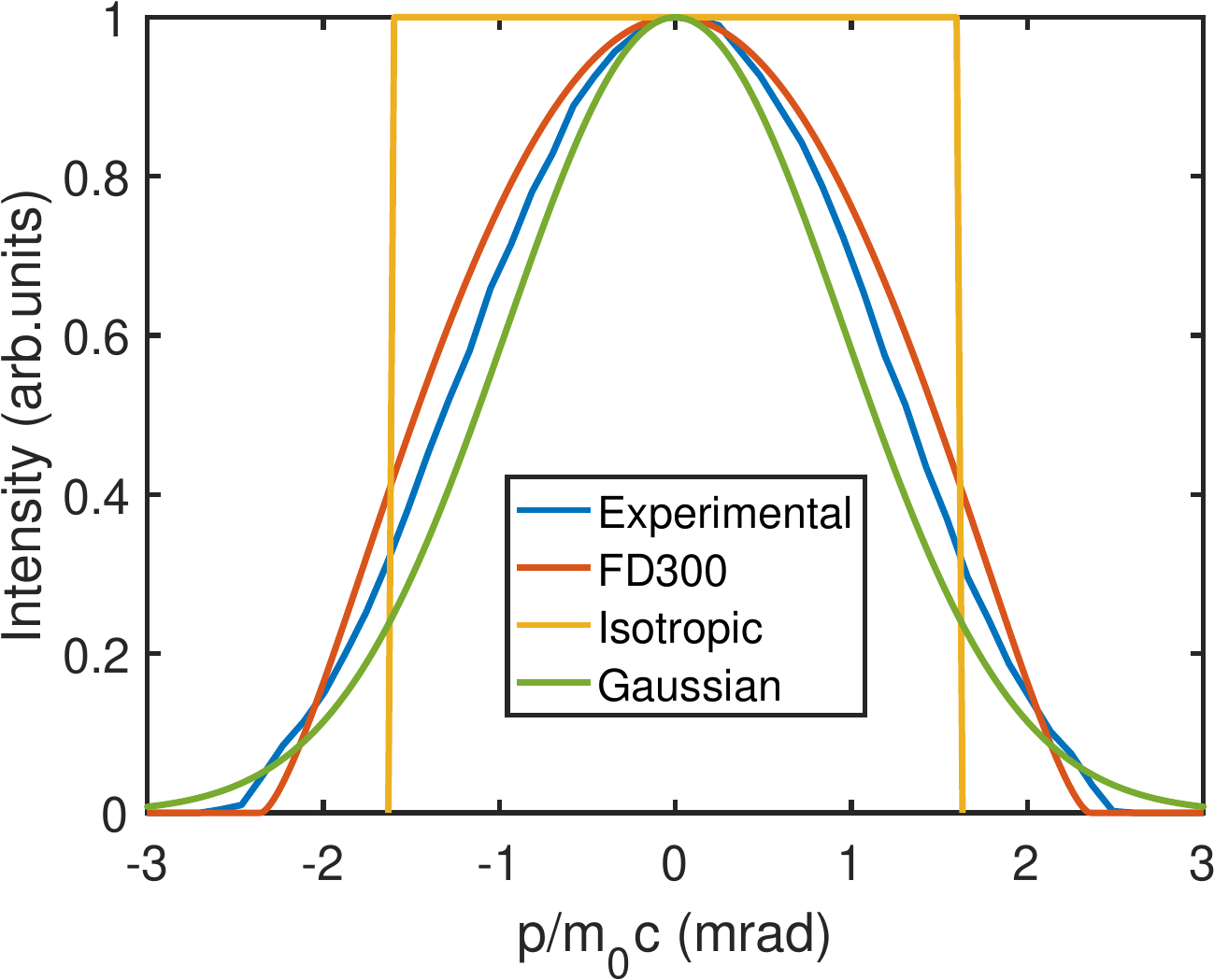}
   \caption{ The measured transverse momentum distribution and different emission models in ASTRA code, which are FD300, isotropic and Gaussian, respectively. }\label{mom_dis}
\end{figure}

With proper imaging condition and reasonable bunch charge, images can be taken for cathode transverse momentum measurement. A typical image is presented in Fig.\;\ref{typical_pic}. In this case, the rms size at the observation screen is around 0.27~mm, corresponding to 0.972 mm mrad/mm for normalized cathode transverse momentum. The distributions of transverse momentum in the horizontal and vertical directions are also displayed in the figure. The obtained momentum spectra is compared with three different emission models in ASTRA, namely FD300, isotropic and Gaussian. FD300 is a typical model for metal \cite{Dowell2009Quantum}. In the model, electrons should follow Fermi-Dirac distribution and occupy all the states below Fermi energy. The density of states is assumed to be constant, leading to the equal probability of electrons' energy. After excitation, the electrons in the material are isotropic in momentum. Only those electrons with longitudinal energy larger than the vacuum barrier can transit into vacuum. As for the isotropic model \cite{flottmann1997note}, the emitted electrons are assumed to have the same energy and the initial direction is isotropic out of the material. The Gaussian model assumes that the transverse momentum of the emitted electrons follows a Gaussian distribution. The corresponding momentum distributions for the three models with the same normalized cathode transverse momentum are compared with the experiment results, as shown in Fig.\;\ref{mom_dis}. Our measured distribution is in between FD300 model and Gaussian model, and very different from isotropic model.

\begin{figure}[!tbp]
   \includegraphics*[width=.8\columnwidth]{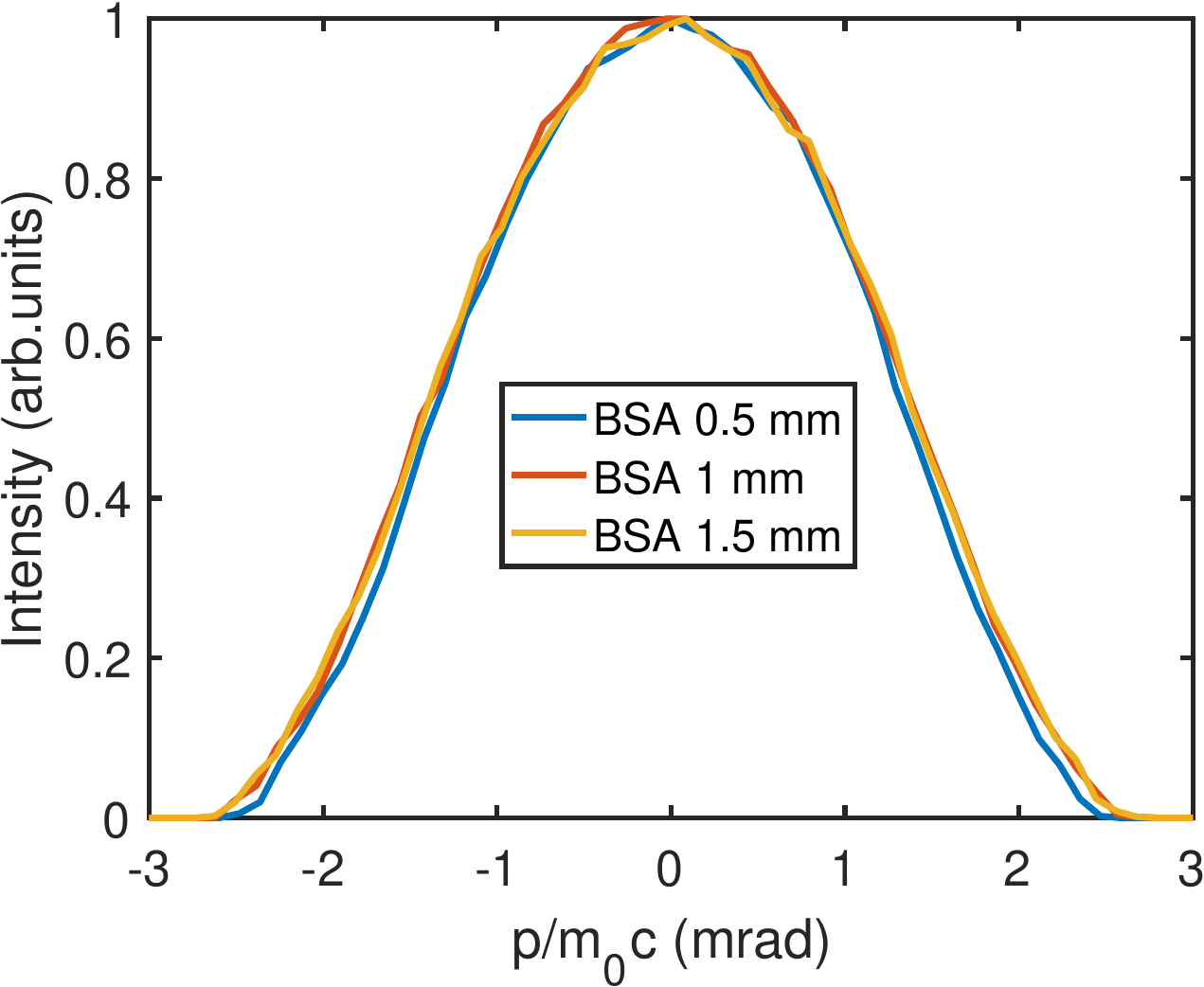}
   \caption{ The measured profiles of transverse momentum with different laser diameters, namely 0.5 mm, 1 mm and 1.5 mm. }\label{profile_bsa}
\end{figure}

The momentum imaging is insensitive against the laser size. Measurements have been taken with three different laser sizes, as shown in Fig.\;\ref{profile_bsa}. The obtained momentum profiles overlap with each other and the corresponding normalized cathode transverse momentum is consistent. This represents an additional confirmation that $M_{11}$ is actually set to zero.

\begin{figure}[!tbp]
	\centering
    \includegraphics*[width=.8\columnwidth]{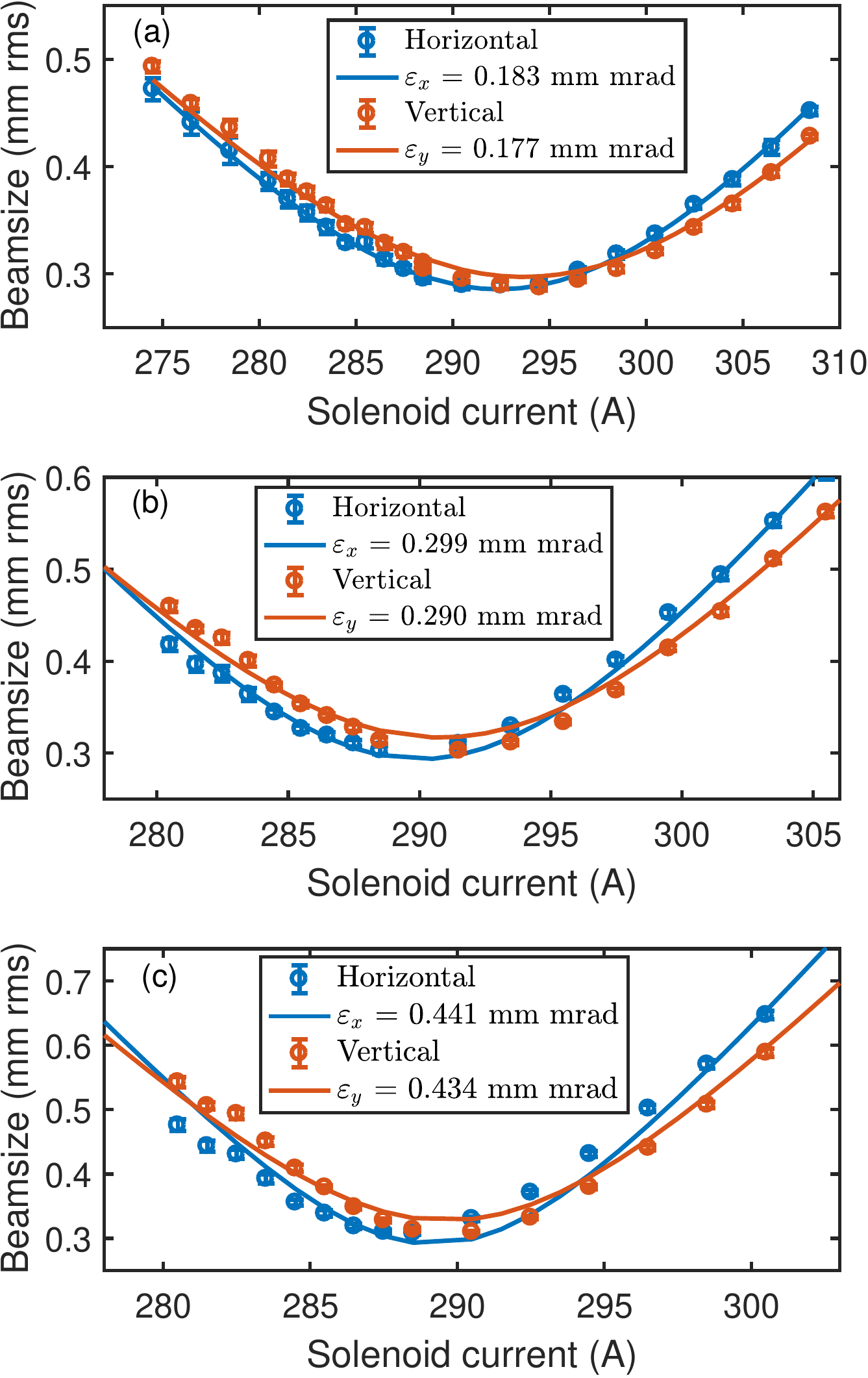}
	\caption{The measurement results of solenoid scans. (a), (b) and (c) refer to the cases with laser diameter 0.6 mm, 1.0 mm and 1.5 mm.} \label{sol_results}
\end{figure}

\begin{figure}[!tbp]
   \includegraphics*[width=.8\columnwidth]{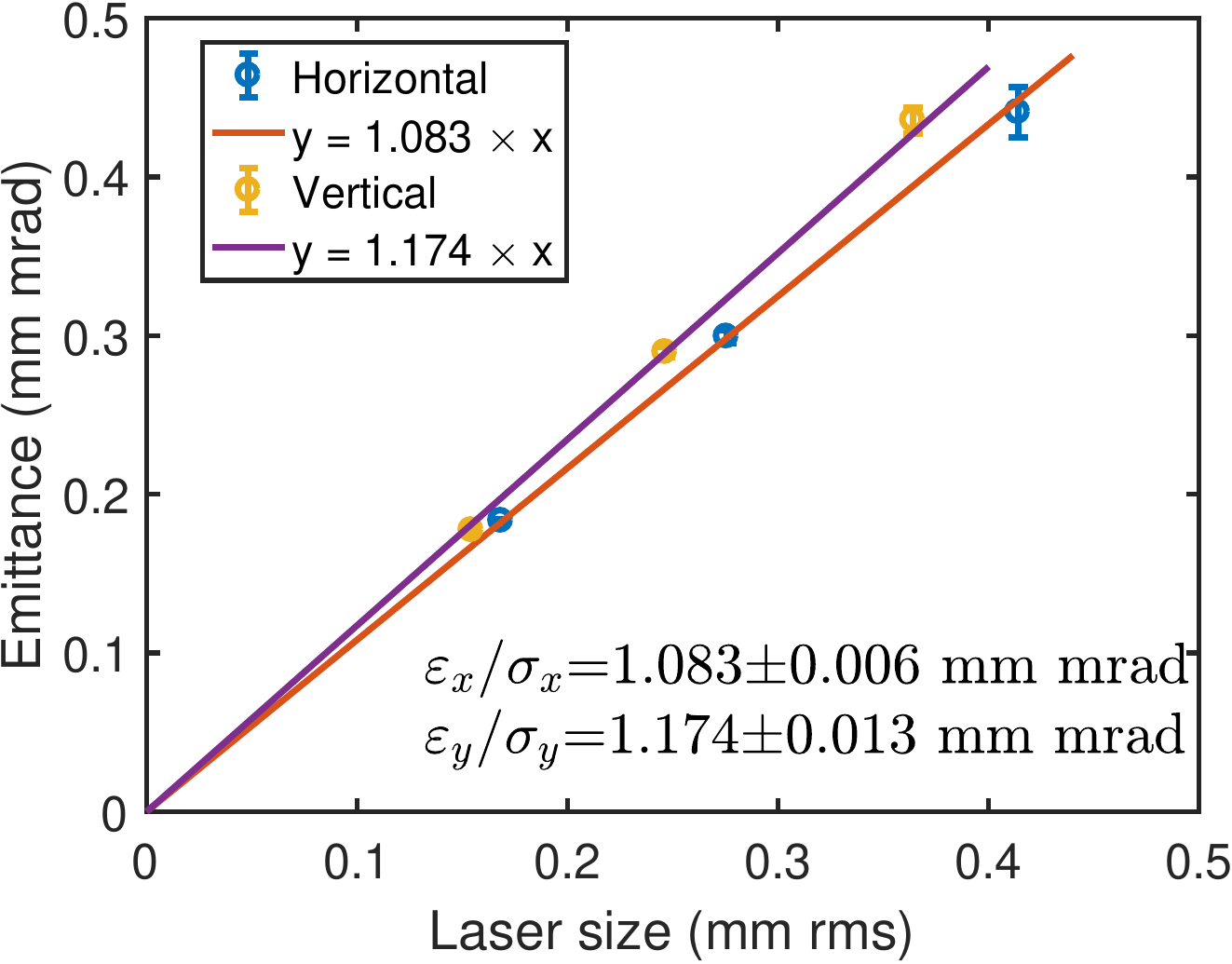}
   \caption{ Cathode transverse momentum is obtained from the linear fit of the beam emittance against the laser spot size. }\label{sol_scan}
\end{figure}

\begin{table}[!tbp]
  \caption{Comparison between the imaging method and traditional solenoid scan}\label{com}
\begin{threeparttable}
\begin{tabular}{ccc}
  \toprule[1pt]
  Method & Horizontal & Vertical \\
         & (mm mrad/mm) & (mm mrad/mm) \\
  \midrule[0.7pt]
  Imaging & 1.097 $\pm$ 0.015 & 1.135 $\pm$ 0.012 \\
  Solenoid scan & 1.083 $\pm$ 0.006 & 1.174 $\pm$ 0.013 \\
  \bottomrule[1pt]
\end{tabular}
\end{threeparttable}
\end{table}

\begin{figure}[!h]
   \includegraphics*[width=.9\columnwidth]{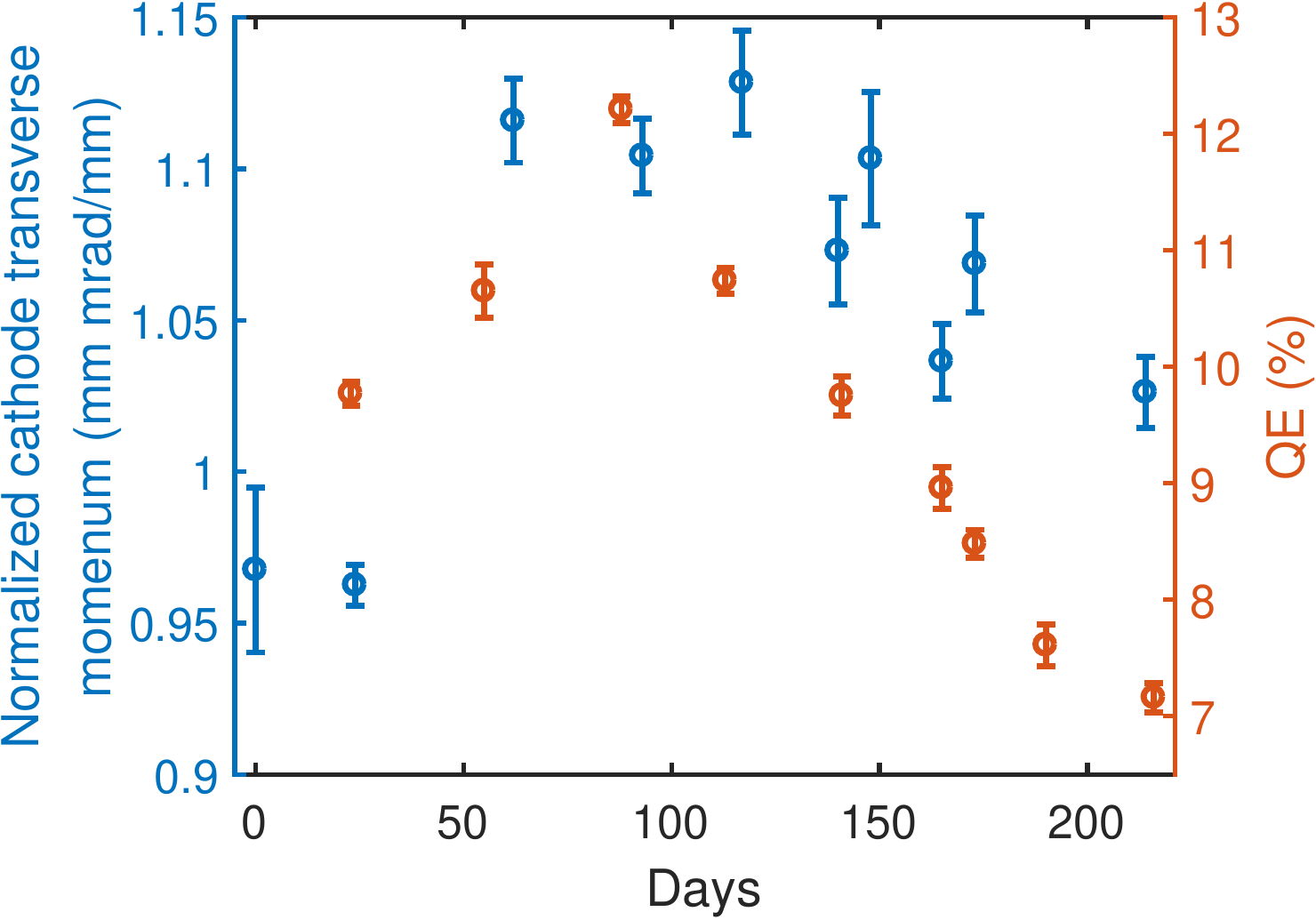}
   \caption{ The evolution of thermal emittance and QE for the Cs$ _2$Te photocathode $\#$661.1 in the PITZ beam line. The blue data points refer to the normalized cathode transverse momentum and the corresponding vertical axis is on the left. The red data points refer to the QE and the corresponding vertical axis is on the right. The emission field at MMMG phase is roughly 40 MV/m for the measurements and the laser wavelength is 257 nm. }\label{emit_and_qe_evo}
\end{figure}
To cross-check the proposed method and the reliability of the value of $M_{12}$ from simulation, the thermal emittance is measured with the more traditional method of the solenoid scan, where the value for this quantity is obtained by measuring and fitting beam sizes at the screen for different solenoid currents. Independent scans were carried out for three different laser diameters, 0.6 mm, 1 mm and 1.5 mm. The corresponding measurement results and fittings are displayed in Fig.\;\ref{sol_results}. The measured thermal emittances versus the laser spots at the cathodes are then fit by a straight line whose slope gives the value of normalized cathode transverse momentum, as shown in Fig.\;\ref{sol_scan}. The results of the two methods are compared in Table.\;\ref{com}. The excellent agreement between the two methods strongly confirms the validity of the proposed method, and positively confirms the reliability of the $M_{12}$ value provided by the simulations. A major advantage of the proposed imaging method with respect to the solenoid method is the capability of measuring the cathode transverse momentum from a single beam size measurement. The capability for fast measurements enables routine measurements of the cathode thermal emittance.

\section{Application}
Routine measurements of the normalized transverse momentum and QE were performed for the Cs$ _2$Te cathode $\#$661.1 at PITZ over a 200-day period. To our knowledge, this is the first time that these two important cathode parameters are simultaneously measured over such a long time period. All the data were taken at the cathode center with the same RF condition and the results are shown in Fig.\;\ref{emit_and_qe_evo}. Day 0 refers to the starting day of the measurements and it is roughly one month after the fresh cathode insertion into the gun. An increase has been observed in both the QE and normalized cathode transverse momentum during the first three months. After that, the QE degraded continously, accompanied by a similar decrease of the cathode transverse momentum. Qualitatively, the similar evolution of cathode QE and transverse momentum is consistent with variations of the effective electron affinity. Further studies are required to understand the degradation process.

\section{Conclusion}

Cathode transverse momentum imaging is proposed as a new method to measure thermal emittance in a photoelectron RF gun. The transverse momentum of the electrons at the cathode is imaged on an observation screen by proper setting of a solenoid. The method, in principle, can also be applied to a thermionic or field emission gun. Contrary to the standard solenoid scan method, the presented technique is based on a single shot measurement that allows for an strongly improved time efficiency. The feasibility of the method has been verified by simulations and demonstrated by a proof of principle experiment. The normalized cathode transverse momentum of a Cs$ _2$Te photocathode in the PITZ gun was measured as an example. The systematic errors of this method due to RF emittance and quadrupole field error in the gun solenoid are discussed, especially in extremely small normalized cathode transverse momentum ($<$ 0.1 mm mrad/mm), and methods are found to suppress errors within an acceptable level. The proposed method was also used to take a long period measurement of the thermal emittance and quantum efficiency of a Cs$ _2$Te photocathode at PITZ and the related results are presented. Such a time efficient measurement technique will enable, for example, more experimental studies on cathode transverse momentum in high gradient guns, which will help to improve the photoemission modeling in high brightness photoinjectors.

\bibliography{reference}

\begin{thebibliography}{23}
\expandafter\ifx\csname natexlab\endcsname\relax\def\natexlab#1{#1}\fi
\expandafter\ifx\csname bibnamefont\endcsname\relax
  \def\bibnamefont#1{#1}\fi
\expandafter\ifx\csname bibfnamefont\endcsname\relax
  \def\bibfnamefont#1{#1}\fi
\expandafter\ifx\csname citenamefont\endcsname\relax
  \def\citenamefont#1{#1}\fi
\expandafter\ifx\csname url\endcsname\relax
  \def\url#1{\texttt{#1}}\fi
\expandafter\ifx\csname urlprefix\endcsname\relax\def\urlprefix{URL }\fi
\providecommand{\bibinfo}[2]{#2}
\providecommand{\eprint}[2][]{\url{#2}}

\bibitem[{\citenamefont{Lessner et~al.}(2016)\citenamefont{Lessner, Wang, and
  Musumeci}}]{lessner2016report}
\bibinfo{author}{\bibfnamefont{E.}~\bibnamefont{Lessner}},
  \bibinfo{author}{\bibfnamefont{X.}~\bibnamefont{Wang}}, \bibnamefont{and}
  \bibinfo{author}{\bibfnamefont{P.}~\bibnamefont{Musumeci}},
  \bibinfo{journal}{SLAC National Accelerator Laboratory}
  (\bibinfo{year}{2016}).

\bibitem[{\citenamefont{Musumeci et~al.}(2018)\citenamefont{Musumeci, Navarro,
  Rosenzweig, Cultrera, Bazarov, Maxson, Karkare, and
  Padmore}}]{musumeci2018advances}
\bibinfo{author}{\bibfnamefont{P.}~\bibnamefont{Musumeci}},
  \bibinfo{author}{\bibfnamefont{J.~G.} \bibnamefont{Navarro}},
  \bibinfo{author}{\bibfnamefont{J.}~\bibnamefont{Rosenzweig}},
  \bibinfo{author}{\bibfnamefont{L.}~\bibnamefont{Cultrera}},
  \bibinfo{author}{\bibfnamefont{I.}~\bibnamefont{Bazarov}},
  \bibinfo{author}{\bibfnamefont{J.}~\bibnamefont{Maxson}},
  \bibinfo{author}{\bibfnamefont{S.}~\bibnamefont{Karkare}}, \bibnamefont{and}
  \bibinfo{author}{\bibfnamefont{H.}~\bibnamefont{Padmore}},
  \bibinfo{journal}{Nuclear Instruments and Methods in Physics Research Section
  A: Accelerators, Spectrometers, Detectors and Associated Equipment}
  \textbf{\bibinfo{volume}{907}}, \bibinfo{pages}{209} (\bibinfo{year}{2018}).

\bibitem[{\citenamefont{Dowell et~al.}(2010)\citenamefont{Dowell, Bazarov,
  Dunham, Harkay, Hernandez-Garcia, Legg, Padmore, Rao, Smedley, and
  Wan}}]{dowell2010cathode}
\bibinfo{author}{\bibfnamefont{D.}~\bibnamefont{Dowell}},
  \bibinfo{author}{\bibfnamefont{I.}~\bibnamefont{Bazarov}},
  \bibinfo{author}{\bibfnamefont{B.}~\bibnamefont{Dunham}},
  \bibinfo{author}{\bibfnamefont{K.}~\bibnamefont{Harkay}},
  \bibinfo{author}{\bibfnamefont{C.}~\bibnamefont{Hernandez-Garcia}},
  \bibinfo{author}{\bibfnamefont{R.}~\bibnamefont{Legg}},
  \bibinfo{author}{\bibfnamefont{H.}~\bibnamefont{Padmore}},
  \bibinfo{author}{\bibfnamefont{T.}~\bibnamefont{Rao}},
  \bibinfo{author}{\bibfnamefont{J.}~\bibnamefont{Smedley}}, \bibnamefont{and}
  \bibinfo{author}{\bibfnamefont{W.}~\bibnamefont{Wan}},
  \bibinfo{journal}{Nuclear Instruments and Methods in Physics Research Section
  A: Accelerators, Spectrometers, Detectors and Associated Equipment}
  \textbf{\bibinfo{volume}{622}}, \bibinfo{pages}{685} (\bibinfo{year}{2010}).

\bibitem[{\citenamefont{Krasilnikov et~al.}(2012)\citenamefont{Krasilnikov,
  Stephan, Asova, Grabosch, Gro{\ss}, Hakobyan, Isaev, Ivanisenko, Jachmann,
  Khojoyan et~al.}}]{pitzpaper}
\bibinfo{author}{\bibfnamefont{M.}~\bibnamefont{Krasilnikov}},
  \bibinfo{author}{\bibfnamefont{F.}~\bibnamefont{Stephan}},
  \bibinfo{author}{\bibfnamefont{G.}~\bibnamefont{Asova}},
  \bibinfo{author}{\bibfnamefont{H.-J.} \bibnamefont{Grabosch}},
  \bibinfo{author}{\bibfnamefont{M.}~\bibnamefont{Gro{\ss}}},
  \bibinfo{author}{\bibfnamefont{L.}~\bibnamefont{Hakobyan}},
  \bibinfo{author}{\bibfnamefont{I.}~\bibnamefont{Isaev}},
  \bibinfo{author}{\bibfnamefont{Y.}~\bibnamefont{Ivanisenko}},
  \bibinfo{author}{\bibfnamefont{L.}~\bibnamefont{Jachmann}},
  \bibinfo{author}{\bibfnamefont{M.}~\bibnamefont{Khojoyan}},
  \bibnamefont{et~al.}, \bibinfo{journal}{Physical Review Special
  Topics-Accelerators and Beams} \textbf{\bibinfo{volume}{15}},
  \bibinfo{pages}{100701} (\bibinfo{year}{2012}).

\bibitem[{\citenamefont{Gulliford et~al.}(2015)\citenamefont{Gulliford,
  Bartnik, Bazarov, Dunham, and Cultrera}}]{cedominate}
\bibinfo{author}{\bibfnamefont{C.}~\bibnamefont{Gulliford}},
  \bibinfo{author}{\bibfnamefont{A.}~\bibnamefont{Bartnik}},
  \bibinfo{author}{\bibfnamefont{I.}~\bibnamefont{Bazarov}},
  \bibinfo{author}{\bibfnamefont{B.}~\bibnamefont{Dunham}}, \bibnamefont{and}
  \bibinfo{author}{\bibfnamefont{L.}~\bibnamefont{Cultrera}},
  \bibinfo{journal}{Applied Physics Letters} \textbf{\bibinfo{volume}{106}},
  \bibinfo{pages}{094101} (\bibinfo{year}{2015}).

\bibitem[{\citenamefont{Mammei et~al.}(2013)\citenamefont{Mammei, Suleiman,
  Feingold, Adderley, Clark, Covert, Grames, Hansknecht, Machie, Poelker
  et~al.}}]{Charge2013}
\bibinfo{author}{\bibfnamefont{R.~R.} \bibnamefont{Mammei}},
  \bibinfo{author}{\bibfnamefont{R.}~\bibnamefont{Suleiman}},
  \bibinfo{author}{\bibfnamefont{J.}~\bibnamefont{Feingold}},
  \bibinfo{author}{\bibfnamefont{P.~A.} \bibnamefont{Adderley}},
  \bibinfo{author}{\bibfnamefont{J.}~\bibnamefont{Clark}},
  \bibinfo{author}{\bibfnamefont{S.}~\bibnamefont{Covert}},
  \bibinfo{author}{\bibfnamefont{J.}~\bibnamefont{Grames}},
  \bibinfo{author}{\bibfnamefont{J.}~\bibnamefont{Hansknecht}},
  \bibinfo{author}{\bibfnamefont{D.}~\bibnamefont{Machie}},
  \bibinfo{author}{\bibfnamefont{M.}~\bibnamefont{Poelker}},
  \bibnamefont{et~al.}, \bibinfo{journal}{Phys. Rev. ST Accel. Beams}
  \textbf{\bibinfo{volume}{16}}, \bibinfo{pages}{033401}
  (\bibinfo{year}{2013}).

\bibitem[{\citenamefont{Le~Pimpec et~al.}(2013)\citenamefont{Le~Pimpec, Milne,
  Hauri, and Ardana-Lamas}}]{le2013quantum}
\bibinfo{author}{\bibfnamefont{F.}~\bibnamefont{Le~Pimpec}},
  \bibinfo{author}{\bibfnamefont{C.}~\bibnamefont{Milne}},
  \bibinfo{author}{\bibfnamefont{C.}~\bibnamefont{Hauri}}, \bibnamefont{and}
  \bibinfo{author}{\bibfnamefont{F.}~\bibnamefont{Ardana-Lamas}},
  \bibinfo{journal}{Applied Physics A} \textbf{\bibinfo{volume}{112}},
  \bibinfo{pages}{647} (\bibinfo{year}{2013}).

\bibitem[{\citenamefont{Filippetto et~al.}(2015)\citenamefont{Filippetto, Qian,
  and Sannibale}}]{Filippetto2015Cesium}
\bibinfo{author}{\bibfnamefont{D.}~\bibnamefont{Filippetto}},
  \bibinfo{author}{\bibfnamefont{H.}~\bibnamefont{Qian}}, \bibnamefont{and}
  \bibinfo{author}{\bibfnamefont{F.}~\bibnamefont{Sannibale}},
  \bibinfo{journal}{Applied Physics Letters} \textbf{\bibinfo{volume}{107}},
  \bibinfo{pages}{042104} (\bibinfo{year}{2015}).

\bibitem[{\citenamefont{Jones et~al.}(2017)\citenamefont{Jones, Scheibler,
  Gorshkov, Terekhov, Militsyn, and Noakes}}]{Jones2017Evolution}
\bibinfo{author}{\bibfnamefont{L.}~\bibnamefont{Jones}},
  \bibinfo{author}{\bibfnamefont{H.}~\bibnamefont{Scheibler}},
  \bibinfo{author}{\bibfnamefont{D.}~\bibnamefont{Gorshkov}},
  \bibinfo{author}{\bibfnamefont{A.}~\bibnamefont{Terekhov}},
  \bibinfo{author}{\bibfnamefont{B.}~\bibnamefont{Militsyn}}, \bibnamefont{and}
  \bibinfo{author}{\bibfnamefont{T.}~\bibnamefont{Noakes}},
  \bibinfo{journal}{Journal of Applied Physics} \textbf{\bibinfo{volume}{121}},
  \bibinfo{pages}{225703} (\bibinfo{year}{2017}).

\bibitem[{\citenamefont{Huang et~al.}(2019)\citenamefont{Huang, Qian, Du,
  Huang, Zhang, and Tang}}]{Huang2019Photoemission}
\bibinfo{author}{\bibfnamefont{P.-W.} \bibnamefont{Huang}},
  \bibinfo{author}{\bibfnamefont{H.}~\bibnamefont{Qian}},
  \bibinfo{author}{\bibfnamefont{Y.}~\bibnamefont{Du}},
  \bibinfo{author}{\bibfnamefont{W.}~\bibnamefont{Huang}},
  \bibinfo{author}{\bibfnamefont{Z.}~\bibnamefont{Zhang}}, \bibnamefont{and}
  \bibinfo{author}{\bibfnamefont{C.}~\bibnamefont{Tang}},
  \bibinfo{journal}{Phys. Rev. Accel. Beams} \textbf{\bibinfo{volume}{22}},
  \bibinfo{pages}{123403} (\bibinfo{year}{2019}).

\bibitem[{\citenamefont{Lee et~al.}(2015)\citenamefont{Lee, Karkare, Cultrera,
  Kim, and Bazarov}}]{lee2015review}
\bibinfo{author}{\bibfnamefont{H.}~\bibnamefont{Lee}},
  \bibinfo{author}{\bibfnamefont{S.}~\bibnamefont{Karkare}},
  \bibinfo{author}{\bibfnamefont{L.}~\bibnamefont{Cultrera}},
  \bibinfo{author}{\bibfnamefont{A.}~\bibnamefont{Kim}}, \bibnamefont{and}
  \bibinfo{author}{\bibfnamefont{I.~V.} \bibnamefont{Bazarov}},
  \bibinfo{journal}{Review of Scientific Instruments}
  \textbf{\bibinfo{volume}{86}}, \bibinfo{pages}{073309}
  (\bibinfo{year}{2015}).

\bibitem[{\citenamefont{Feng et~al.}(2015{\natexlab{a}})\citenamefont{Feng,
  Nasiatka, Wan, Vecchione, and Padmore}}]{feng2015novel}
\bibinfo{author}{\bibfnamefont{J.}~\bibnamefont{Feng}},
  \bibinfo{author}{\bibfnamefont{J.}~\bibnamefont{Nasiatka}},
  \bibinfo{author}{\bibfnamefont{W.}~\bibnamefont{Wan}},
  \bibinfo{author}{\bibfnamefont{T.}~\bibnamefont{Vecchione}},
  \bibnamefont{and} \bibinfo{author}{\bibfnamefont{H.}~\bibnamefont{Padmore}},
  \bibinfo{journal}{Review of Scientific Instruments}
  \textbf{\bibinfo{volume}{86}}, \bibinfo{pages}{015103}
  (\bibinfo{year}{2015}{\natexlab{a}}).

\bibitem[{\citenamefont{Berger et~al.}(2012)\citenamefont{Berger, Rickman, Li,
  Nicholls, and Andreas~Schroeder}}]{Berger2012excited}
\bibinfo{author}{\bibfnamefont{J.~A.} \bibnamefont{Berger}},
  \bibinfo{author}{\bibfnamefont{B.}~\bibnamefont{Rickman}},
  \bibinfo{author}{\bibfnamefont{T.}~\bibnamefont{Li}},
  \bibinfo{author}{\bibfnamefont{A.}~\bibnamefont{Nicholls}}, \bibnamefont{and}
  \bibinfo{author}{\bibfnamefont{W.}~\bibnamefont{Andreas~Schroeder}},
  \bibinfo{journal}{Applied Physics Letters} \textbf{\bibinfo{volume}{101}},
  \bibinfo{pages}{194103} (\bibinfo{year}{2012}).

\bibitem[{\citenamefont{Kim}(1989)}]{kim1989rf}
\bibinfo{author}{\bibfnamefont{K.-J.} \bibnamefont{Kim}},
  \bibinfo{journal}{Nuclear Instruments and Methods in Physics Research Section
  A: Accelerators, Spectrometers, Detectors and Associated Equipment}
  \textbf{\bibinfo{volume}{275}}, \bibinfo{pages}{201} (\bibinfo{year}{1989}).

\bibitem[{\citenamefont{Floettmann}(2007)}]{astra}
\bibinfo{author}{\bibfnamefont{K.}~\bibnamefont{Floettmann}},
  \bibinfo{journal}{ASTRAURL http://www. desy. de/mpyflo}
  (\bibinfo{year}{2007}).

\bibitem[{\citenamefont{Niemczyk et~al.}(2019)\citenamefont{Niemczyk,
  Boonpornprasert, Chen, Good, Gro{\ss}, Huck, Isaev, Kalantaryan, Koschitzki,
  Krasilnikov et~al.}}]{niemczyk2019comparison}
\bibinfo{author}{\bibfnamefont{R.}~\bibnamefont{Niemczyk}},
  \bibinfo{author}{\bibfnamefont{P.}~\bibnamefont{Boonpornprasert}},
  \bibinfo{author}{\bibfnamefont{Y.}~\bibnamefont{Chen}},
  \bibinfo{author}{\bibfnamefont{J.}~\bibnamefont{Good}},
  \bibinfo{author}{\bibfnamefont{M.}~\bibnamefont{Gro{\ss}}},
  \bibinfo{author}{\bibfnamefont{H.}~\bibnamefont{Huck}},
  \bibinfo{author}{\bibfnamefont{I.}~\bibnamefont{Isaev}},
  \bibinfo{author}{\bibfnamefont{D.}~\bibnamefont{Kalantaryan}},
  \bibinfo{author}{\bibfnamefont{C.}~\bibnamefont{Koschitzki}},
  \bibinfo{author}{\bibfnamefont{M.}~\bibnamefont{Krasilnikov}},
  \bibnamefont{et~al.}, in \emph{\bibinfo{booktitle}{7$ ^{th}$ Int. Beam
  Instrumentation Conf.(IBIC'18), Shanghai, China, 09-13 September 2018}}
  (\bibinfo{organization}{JACOW Publishing, Geneva, Switzerland},
  \bibinfo{year}{2019}), pp. \bibinfo{pages}{438--440}.

\bibitem[{\citenamefont{Feng et~al.}(2015{\natexlab{b}})\citenamefont{Feng,
  Nasiatka, Wan, Karkare, Smedley, and Padmore}}]{feng2015thermal}
\bibinfo{author}{\bibfnamefont{J.}~\bibnamefont{Feng}},
  \bibinfo{author}{\bibfnamefont{J.}~\bibnamefont{Nasiatka}},
  \bibinfo{author}{\bibfnamefont{W.}~\bibnamefont{Wan}},
  \bibinfo{author}{\bibfnamefont{S.}~\bibnamefont{Karkare}},
  \bibinfo{author}{\bibfnamefont{J.}~\bibnamefont{Smedley}}, \bibnamefont{and}
  \bibinfo{author}{\bibfnamefont{H.~A.} \bibnamefont{Padmore}},
  \bibinfo{journal}{Applied Physics Letters} \textbf{\bibinfo{volume}{107}},
  \bibinfo{pages}{134101} (\bibinfo{year}{2015}{\natexlab{b}}).

\bibitem[{\citenamefont{Krasilnikov et~al.}(2018)}]{Krasilnikov}
\bibinfo{author}{\bibfnamefont{M.}~\bibnamefont{Krasilnikov}}
  \bibnamefont{et~al.}, in \emph{\bibinfo{booktitle}{Proc. of International
  Free Electron Laser Conference (FEL'17), Santa Fe, NM, USA, August 20-25,
  2017}} (\bibinfo{publisher}{JACoW}, \bibinfo{address}{Geneva, Switzerland},
  \bibinfo{year}{2018}), no.~\bibinfo{number}{38} in
  \bibinfo{series}{International Free Electron Laser Conference}, pp.
  \bibinfo{pages}{429--431}, ISBN \bibinfo{isbn}{978-3-95450-179-3}.

\bibitem[{\citenamefont{Dowell et~al.}(2018)\citenamefont{Dowell, Zhou, and
  Schmerge}}]{Dowell2018solenoid}
\bibinfo{author}{\bibfnamefont{D.~H.} \bibnamefont{Dowell}},
  \bibinfo{author}{\bibfnamefont{F.}~\bibnamefont{Zhou}}, \bibnamefont{and}
  \bibinfo{author}{\bibfnamefont{J.}~\bibnamefont{Schmerge}},
  \bibinfo{journal}{Phys. Rev. Accel. Beams} \textbf{\bibinfo{volume}{21}},
  \bibinfo{pages}{010101} (\bibinfo{year}{2018}).

\bibitem[{\citenamefont{Zheng et~al.}(2018)\citenamefont{Zheng, Shao, Du,
  Power, Wisniewski, Liu, Whiteford, Conde, Doran, Jing
  et~al.}}]{Zheng2018overestimation}
\bibinfo{author}{\bibfnamefont{L.}~\bibnamefont{Zheng}},
  \bibinfo{author}{\bibfnamefont{J.}~\bibnamefont{Shao}},
  \bibinfo{author}{\bibfnamefont{Y.}~\bibnamefont{Du}},
  \bibinfo{author}{\bibfnamefont{J.~G.} \bibnamefont{Power}},
  \bibinfo{author}{\bibfnamefont{E.~E.} \bibnamefont{Wisniewski}},
  \bibinfo{author}{\bibfnamefont{W.}~\bibnamefont{Liu}},
  \bibinfo{author}{\bibfnamefont{C.~E.} \bibnamefont{Whiteford}},
  \bibinfo{author}{\bibfnamefont{M.}~\bibnamefont{Conde}},
  \bibinfo{author}{\bibfnamefont{S.}~\bibnamefont{Doran}},
  \bibinfo{author}{\bibfnamefont{C.}~\bibnamefont{Jing}}, \bibnamefont{et~al.},
  \bibinfo{journal}{Phys. Rev. Accel. Beams} \textbf{\bibinfo{volume}{21}},
  \bibinfo{pages}{122803} (\bibinfo{year}{2018}).

\bibitem[{\citenamefont{Zheng et~al.}(2019)\citenamefont{Zheng, Shao, Du,
  Power, Wisniewski, Liu, Whiteford, Conde, Doran, Jing
  et~al.}}]{Zheng2019Experimental}
\bibinfo{author}{\bibfnamefont{L.}~\bibnamefont{Zheng}},
  \bibinfo{author}{\bibfnamefont{J.}~\bibnamefont{Shao}},
  \bibinfo{author}{\bibfnamefont{Y.}~\bibnamefont{Du}},
  \bibinfo{author}{\bibfnamefont{J.~G.} \bibnamefont{Power}},
  \bibinfo{author}{\bibfnamefont{E.~E.} \bibnamefont{Wisniewski}},
  \bibinfo{author}{\bibfnamefont{W.}~\bibnamefont{Liu}},
  \bibinfo{author}{\bibfnamefont{C.~E.} \bibnamefont{Whiteford}},
  \bibinfo{author}{\bibfnamefont{M.}~\bibnamefont{Conde}},
  \bibinfo{author}{\bibfnamefont{S.}~\bibnamefont{Doran}},
  \bibinfo{author}{\bibfnamefont{C.}~\bibnamefont{Jing}}, \bibnamefont{et~al.},
  \bibinfo{journal}{Phys. Rev. Accel. Beams} \textbf{\bibinfo{volume}{22}},
  \bibinfo{pages}{072805} (\bibinfo{year}{2019}).

\bibitem[{\citenamefont{Dowell and Schmerge}(2009)}]{Dowell2009Quantum}
\bibinfo{author}{\bibfnamefont{D.~H.} \bibnamefont{Dowell}} \bibnamefont{and}
  \bibinfo{author}{\bibfnamefont{J.~F.} \bibnamefont{Schmerge}},
  \bibinfo{journal}{Phys. Rev. ST Accel. Beams} \textbf{\bibinfo{volume}{12}},
  \bibinfo{pages}{074201} (\bibinfo{year}{2009}),
  \urlprefix\url{https://link.aps.org/doi/10.1103/PhysRevSTAB.12.074201}.

\bibitem[{\citenamefont{Fl{\"o}ttmann}(1997)}]{flottmann1997note}
\bibinfo{author}{\bibfnamefont{K.}~\bibnamefont{Fl{\"o}ttmann}},
  \bibinfo{type}{Tech. Rep.}, \bibinfo{institution}{SCAN-9708052}
  (\bibinfo{year}{1997}).

\end{thebibliography}
\end{document}